\documentclass[structabstract,bibyear,longauth]{aa}
\usepackage{graphicx}
\usepackage{txfonts}
\begin{document}

\title{The Lyman Continuum escape fraction of faint galaxies at $z\sim 3.3$
in the CANDELS/GOODS-North, EGS, and COSMOS fields with LBC\thanks{Based on
observations made at the Large Binocular Telescope
(LBT) at Mt. Graham (Arizona, USA).}}

\author{A. Grazian\inst{1}
\and E. Giallongo\inst{1}
\and D. Paris\inst{1}
\and K. Boutsia\inst{2}
\and M. Dickinson\inst{3}
\and P. Santini\inst{1}
\and R. A. Windhorst\inst{4}
\and R. A. Jansen\inst{4}
\and S. H. Cohen\inst{4}
\and T. A. Ashcraft\inst{4}
\and C. Scarlata\inst{5}
\and M. J. Rutkowski\inst{6}
\and E. Vanzella\inst{7}
\and F. Cusano\inst{7}
\and S. Cristiani\inst{8}
\and M. Giavalisco\inst{9}
\and H. C. Ferguson\inst{10}
\and A. Koekemoer\inst{10}
\and N. A. Grogin\inst{10}
\and M. Castellano\inst{1}
\and F. Fiore\inst{1}
\and A. Fontana\inst{1}
\and F. Marchi\inst{1}
\and F. Pedichini\inst{1}
\and L. Pentericci\inst{1}
\and R. Amor\'in\inst{11,12}
\and G. Barro\inst{13}
\and A. Bonchi\inst{14,1}
\and A. Bongiorno\inst{1}
\and S. M. Faber\inst{15}
\and M. Fumana\inst{16}
\and A. Galametz\inst{17}
\and L. Guaita\inst{18,1}
\and D. D. Kocevski\inst{19}
\and E. Merlin\inst{1}
\and M. Nonino\inst{8}
\and R. W. O'Connell\inst{20}
\and S. Pilo\inst{1}
\and R. E. Ryan\inst{10}
\and E. Sani\inst{18}
\and R. Speziali\inst{1}
\and V. Testa\inst{1}
\and B. Weiner\inst{21}
\and H. Yan\inst{22}
}

\offprints{A. Grazian, \email{andrea.grazian@oa-roma.inaf.it}}

\institute{INAF--Osservatorio Astronomico di Roma, Via Frascati 33,
I-00078, Monte Porzio Catone, Italy
\and
Carnegie Observatories, Colina El Pino, Casilla 601 La Serena, Chile 
\and
National Optical Astronomy Observatory, 950 North Cherry Ave,
Tucson, AZ 85719, USA
\and
School of Earth and Space Exploration, Arizona State University, Tempe,
AZ 85287-1404, USA
\and
Minnesota Institute for Astrophysics, University of Minnesota,
116 Church Street SE, Minneapolis, MN 55455, USA
\and
Stockholm University, Alba Nova SCFAB, SE-106 91 Stockholm, Sweden
\and
INAF--Osservatorio Astronomico di Bologna, via Ranzani, 1, I-40127,
Bologna, Italy
\and
INAF--Osservatorio Astronomico di Trieste, Via G.B. Tiepolo, 11
I-34143, Trieste, Italy
\and
Astronomy Department, University of Massachusetts, Amherst, MA 01003, USA
\and
Space Telescope Science Institute, 3700 San Martin Drive, Baltimore,
MD 21218, USA
\and
Cavendish Laboratory, University of Cambridge, 19 J.J. Thomson Ave.,
Cambridge CB30HE, UK
\and
Kavli Institute of Cosmology
c/o Institute of Astronomy,
Madingley Road,
Cambridge CB3 0HA, UK
\and
Department of Astronomy and Astrophysics,
University of California Berkeley
501 Campbell Hall
Berkeley, CA 94720, USA
\and
Agenzia Spaziale Italiana Science Data Center, Via del Politecnico
snc, I-00133, Roma, Italy
\and
UCO/Lick Observatory
1156 High Street
Santa Cruz, CA 95064, USA
\and
INAF--IASF, via Bassini 15, I-20133, Milano, Italy
\and
Max Planck Institute for extraterrestrial Physics,
Giessenbachstrasse 1
D-85748 Garching, Bayern,
Deutschland
\and
ESO Vitacura,
Alonso de Cordova 3107,
Vitacura, Casilla 19001,
Santiago de Chile, Chile
\and
University of Kentucky,
600 Rose Street,
Lexington, KY 40508, USA
\and
Department of Astronomy, University of Virginia, Charlottesville,
VA 22904-4325, USA
\and
Department of Astronomy and Steward Observatory, University of Arizona,
Tucson, AZ 85719, USA
\and
Department of Physics and Astronomy
University of Missouri
Columbia, MO 65211, USA
}

\date{Received Month day, year; accepted Month day, year}

\authorrunning{Grazian et al.}
\titlerunning{The Lyman continuum escape fraction of galaxies at z=3.3}

  \abstract
{
The reionization of the Universe is one of the most important
topics of present day astrophysical research. The most
plausible candidates for the reionization process are star-forming
galaxies, which according to the predictions of the majority of the
theoretical and semi-analytical models should dominate the \ion{H}{I}
ionizing background at $z\gtrsim 3$.
}
{
We aim at measuring the Lyman continuum escape fraction, which is
one of the key parameters to compute the contribution of star-forming
galaxies to the UV background. It provides the ratio between the photons
produced at $\lambda\le 912$ {\AA} rest-frame and those which are able
to reach the CGM/IGM, not being absorbed by the neutral hydrogen or by
the dust of the galaxy's ISM.
}
{
We have used ultra-deep U-band imaging ($U=30.2mag$ at 1$\sigma$)
by LBC/LBT in the
CANDELS/GOODS-North field, as well as deep imaging in COSMOS and EGS fields,
in order to estimate the Lyman continuum escape fraction of 69 star-forming
galaxies with secure spectroscopic redshifts
at $3.27\le z\le 3.40$ to faint magnitude limits
($L=0.2L^*$, or equivalently $M_{1500}\sim -19$). The narrow
redshift range implies that the LBC U-band filter samples exclusively the
$\lambda\le 912$ {\AA} rest-frame wavelengths.
}
{
We have
measured through stacks a stringent upper limit (<1.7\% at $1\sigma$) for the
relative escape fraction of \ion{H}{I}
ionizing photons from bright galaxies ($L>L^*$), while for the faint
population ($L=0.2L^*$) the limit to the escape fraction is
$\lesssim 10\%$. We have computed the contribution of star-forming
galaxies to the observed UV background at $z\sim 3$ and
we have found that it is not enough to keep the Universe ionized at
these redshifts, unless their escape fraction
increases significantly ($\ge 10\%$) at low luminosities ($M_{1500}\ge -19$).
}
{
We compare our results on the Lyman continuum escape fraction of
high-z galaxies with recent estimates in the literature and discuss
future prospects to shed light on the end of the Dark Ages. In the
future, strong gravitational lensing will be fundamental to measure
the Lyman continuum escape fraction down to faint magnitudes ($M_{1500}\sim -16$)
which are inaccessible with the present instrumentation on blank fields.
These results will be important in order to quantify the role of
faint galaxies to the reionization budget.
}


   \maketitle
%

\section{Introduction}

The reionization of the Universe, related to the end of the so-called
Dark Ages, is now located in the redshift interval $z=6.0-8.8$. The
lower limit is derived from observations of the Gunn-Peterson effect
in luminous $z>6$ QSO spectra (\cite{fan06}), while the most recent
upper limit, $z<8.8$, comes from measurements of the Thomson optical
depth $\tau_e=0.055\pm 0.009$ in the CMB polarization map by Planck
(Planck collaboration 2016). These limits are consistent with a
rapid \ion{H}{I} reionization at $z_{reion}=7.8\pm 1.0$.

These main results have been corroborated by a number of
confirmations: the so-called kinetic Sunyaev-Zel'dovich (kSZ) effect
measured by the South Pole Telescope (SPT, \cite{zahn12}), when
combined with the recent Planck 2016 maps gives a stringent limit of
$\Delta z_{reion}<2.8$ to the duration of reionization. Moreover, the
redshift evolution of the luminosity function (\cite{konno14}) of
Lyman-$\alpha$ emitters (LAEs) and the sudden drop observed at $z\sim
7$ in the fraction of Lyman-break galaxies (LBGs) with strong
Lyman-$\alpha$ in emission (e.g.
\cite{fontana10,pentericci11,ono12,treu12,pentericci14}, Schenker et
al. 2014) are indicating that the Universe is becoming more neutral at
$z>7$. All these observations are favoring a late, and possibly
inhomogeneous, reionization process.

While we have now significant information on the timing of the
reionization process, we are still looking for the sources providing
the bulk of the \ion{H}{I} ionizing photons. Obvious candidates have been
searched so far among high-redshift star-forming galaxies (SFGs)
and/or Active Galactic Nuclei (AGNs).

At high redshift, bright QSOs ($M_{1450}\le -27$) are quite rare
as their space density declines rapidly. As such, their ionizing
emissivity is quite low, and they are not enough to keep the
intergalactic medium (IGM) ionized at $z>3$ (Haardt \& Madau 1996,
\cite{cowie09}). However Glikman et al. (2011) and \cite{giallongo15}
have recently found a significant population of fainter ($-25\le
M_{1450}\le -19$) AGNs at $z\ge 4$, which could possibly provide the
whole ionizing emissivity at high-z (\cite{madau15}).

In the last 15 years much effort has been given to high-z star-forming
galaxies as emitters of ionizing radiation. This choice has been
driven by four considerations: 1-early WMAP results, heavily affected by
dust polarization of the Milky Way, indicated a much earlier
reionization epoch, around $z\ge 12$ (\cite{wmap}); 2-the luminosity
function of galaxies is gradually steepening with redshifts at $z\ge
4$ (\cite{finkelstein15,bouwens15}); 3-numerous galaxies exist even
at very faint absolute magnitudes ($M_{1500}\sim -12$) at $z\ge 6$
(\cite{livermore16}); 4-at high redshifts galaxies could be efficient
producers of ionizing photons (\cite{bouwens16}). Their possible
contribution crucially depends, however, on their ability to carve
ionizing bubbles in their neighborhood, which is quantified by the
Lyman Continuum (LyC) escape fraction parameter.

At $z\sim 3-4.5$, the highest redshift where the IGM absorption does not
prevent this kind of measure, we still do not have a reliable estimate
of the average escape fraction from the star-forming galaxy
population. A few claimed detections
(\cite{steidel01,shapley06,nestor11,mostardi13}) have been called into
question due to possible contamination by low redshift interlopers
(e.g. Vanzella et al. 2012a). Other datasets provide only upper limits
at few percent levels (\cite{giallongo02,grimes09,cowie09,siana10,bridge10},
Vanzella et al. 2010, Boutsia et al. 2011, Leitet et al. 2013, Grazian et
al. 2016, Guaita et al. 2016, Japelj et al. 2017) or marginal
detections (Marchi et al. 2017).

Very recently the situation has improved, especially at very low
redshifts, thanks to observations by the COS instrument onboard
HST. Escape fractions within 6-13\% have been detected in five
galaxies with large [OIII]/[OII] line ratio and high ionizing
photon production efficiency. These ionizers appear compact with high
star formation rate surface density $\Sigma_{SFR}$ (Izotov et
al. 2016a,b, \cite{verhamme16,schaerer16}), resembling properties
which are typical of Green Pea galaxies (Cardamone et al. 2009,
Amorin et al. 2010, Amorin et al. 2012a, Henry
et al. 2015). This key feature of the local ionizers is inducing a
strong feedback on their ISM, through rapid outflows, which may be
physically related to leakage of Lyman continuum radiation, as
suggested e.g. by Heckman et al. (2011), Amorin et al. (2012b),
Borthakur et al. (2014), and
explored further with simulations by Sharma et al. (2016a,b). High
[OIII]/[OII] line ratio seen in low-z blue compact dwarfs (BCDs) and high-z
galaxies may be attributed to many possibilities such as low
metallicity, high ionization parameter, hard ionizing radiation field
and/or the presence of density-bound \ion{H}{II} regions (Stasinska
et al. 2015, Steidel et al. 2016, Nakajima et al. 2016). Similar
results were also found by Leitet et al. (2013), Borthakur et
al. (2014), Leitherer et al. (2016), Bergvall et al. (2016). Three
out of the five galaxies by Izotov et al. (2016b) have been detected
by the Wide-field Infrared Survey Explorer (WISE) at 22 micron,
possibly indicating the presence of warm/hot dust, heated by a buried
starburst or by an AGN. The luminosities of these LyC emitters are
between $M_{UV}=-20.4$ and -21.3, thus around $L^*(z=3)$, hereafter $L^*$.
These five galaxies were selected from a sample of $\sim 10^4$ SDSS galaxies at
$z\sim 0.3$, so it is not clear whether they represent a sparse
minority of the SDSS sample, selected thanks to their peculiar
properties (large [OIII]/[OII] line ratio).

Vanzella et al. (2016) and de Barros et al. (2016) found the first
clear evidence of a LyC emitter at $z\sim 3.2$ free from interloper
contamination. This galaxy, dubbed Ion2, has peculiar properties
similar to the five galaxies by Izotov et al. (2016b) mentioned
above. Ion2 has a line ratio of [OIII]/[OII]$>10$, strong
Lyman-$\alpha$ in emission and compact morphology. Galaxies with
similar properties have been recently discovered and characterized by
Amorin et al. (2017), who showed that they are relatively rare and
good analogs of primeval galaxies at $z>6$.  However the spectral and
physical properties of these objects are rather peculiar and it is not
clear if a significant population with similar properties is common at
high redshifts (Faisst 2016, Khostovan et al. 2016). Recently,
\cite{naidu16} have found 6 LyC candidates out of 1124 galaxies at
$z\sim 2$ in the GOODS fields, confirming the peculiarity of these LyC
emitters.  Interestingly, at least 50\% of these ionizing sources are
confirmed as AGNs, with large escape fraction ($\gtrsim 60\%$).

Given the null result on the brighter SFGs, a crucial
contribution on the reionization process by a faint population
($L<L^*$) at high redshift has been envisaged. This hypothesis has
also been predicted by some theoretical models (e.g. Yajima et al. 2011,
Razoumov \& Sommer-Larsen 2010, but see also models by Gnedin et al.
2008, Ma et al. 2015, Sharma et al. 2016a,b for opposite conclusions).

From an observational perspective, however, there are several pieces
of evidence suggesting that low luminosity galaxies may not contribute
appreciably to the \ion{H}{I} ionizing background.  Among the most
interesting results there is the large scale opacity fluctuations
observed in the Lyman-$\alpha$ forest of high-z QSOs by Becker et
al. (2015). The large variance in the spatial scales of the
transmission of the IGM at $z\sim 5.5-6.0$ has been interpreted as due
to bright and rare sources rather than to a diffuse population of
faint galaxies (e.g. Chardin et al. 2017). This indication has been
strengthen by non detection of LyC flux (escape fraction $<2\%$ at
3$\sigma$) from a large sample of moderately star-forming galaxies at
$z\sim 1$ by \cite{rutkowski16}.

An additional observational evidence is related to the sudden decrease
in the number density of Lyman-$\alpha$ emitters at $z\gtrsim 6$.
According to predictions by \cite{dijkstra16}, the escape of
Lyman-$\alpha$ photons should be connected with the leakage of LyC
radiation, since they are possibly linked to clear sight-lines in
neutral hydrogen. The observed drop in Lyman-$\alpha$ is interpreted
as due to a rapid evolution of the ionization fraction of the
intergalactic medium (IGM) in the Universe from $z=6$ to $z=7-8$
(Fontana et al. 2010, Pentericci et al. 2011, 2014, Schmidt et
al. 2016). Moreover, the decrease in number density is differential,
with a less pronounced and slower drop for brighter galaxies
(\cite{matthee15,santos16}, Vanzella et al. 2014, Oesch et al. 2015,
Stark et al. 2017), which can be described as ``Downsizing
Reionization''. This behavior is commonly interpreted as a signature
of inhomogeneous and patchy reionization (Treu et al. 2012, Pentericci
et al. 2014) where ionizing bubbles expand first in cosmic time around
brighter and more massive galaxies, possibly associated with over-dense
regions (Castellano et al. 2016). The suggested average
``Downsizing'' scenario for the luminosity of the ionizing sources
would possibly imply a minor, if any, contribution of faint galaxies
to the reionization process. To improve our knowledge on Lyman
continuum emitters among the SFGs, we started to analyze a relatively
large sample of galaxies at $z\sim 3$, with specific attention to the
fainter population.

In the COSMOS field, the LyC escape fraction of 45 bright ($L\ge 0.5
L^*$) star-forming galaxies at $z\sim 3.3$ has been studied
by \cite{boutsia11} and \cite{grazian16}.
In a companion paper by \cite{guaita16} we have
enlarged this sample adding 86 bright galaxies in the Extended Chandra
Deep Field South. In the present paper we explore the \ion{H}{I} ionizing
contribution of fainter galaxies in the CANDELS/GOODS-North and EGS
fields with the addition of crucial and unique ultra deep UV imaging
from the LBC imager at the Large Binocular Telescope. We have also
included in the analysis three galaxies affected by strong lensing, with the
aim of probing the possible LyC contribution by intrinsically low
luminosity galaxies down to $L\sim 0.04 L^*$.

This paper is organized as follows. In Sect. 2 we present the dataset,
in Sect. 3 we describe the method adopted, in Sect. 4 we show the
results for individual objects and for the overall sample as a whole,
in Sect. 5 we provide an estimate of the ionizing background produced
by these galaxies, in Sect. 6 we discuss our results and in Sect. 7 we
provide a summary and the conclusions. Throughout the paper we adopt
the $\Lambda$-CDM concordance cosmological model ($H_0 = 70~
km/s/Mpc$, $\Omega_M=0.3$ and $\Omega_\Lambda=0.7$), consistent with recent
CMB measurements (Planck collaboration 2016). All magnitudes
are in the AB system. All the limits to the escape fraction in this paper
are at 1$\sigma$ level, unless we state otherwise.


\section{Data}

The sample used in the present analysis is based on new ultra-deep UV
images obtained with the LBC instrument of spectroscopically confirmed
galaxies in the GOODS-North field coupled with moderately deep LBC
images in the EGS/AEGIS field and additional data on the COSMOS field,
collected from the literature as described in the following.

\subsection{The COSMOS, Q0933+28, and Q1623+26 fields}

We started this project with
deep LBC imaging of the COSMOS, Q0933+28, and Q1623+26 fields in the U and R
bands, which have been used by \cite{boutsia11} and \cite{grazian16} to study
the LyC escape fraction of 11 and 34 star-forming galaxies,
respectively, with spectroscopic redshifts $\sim 3.3$ from the VIMOS
Ultra Deep Survey (VUDS, \cite{vuds}). They constitute a mix of Lyman-$\alpha$
emitters and Lyman-Break galaxies. This starting sample of 45
galaxies has been visually inspected against contamination by
foreground objects and it has been discussed in detail in
\cite{grazian16}.

We added to this starting sample 7 galaxies with spectroscopic
redshifts $3.27<z<3.40$ from \cite{onodera16} in the same COSMOS area
covered by deep LBC data in the U and R bands. All these galaxies have
magnitudes between $R=24.5$ and 26.0, with the exception of a very
faint source with $R\sim 27$, and all the seven galaxies have a flux ratio
between [OIII] and [OII] of $\sim 1.5-3$ and SFR of $\sim 4-70M_{\odot}/yr$.
An additional bright ($R=22.95$) star-forming
galaxy (ID=53167 in Table \ref{table:gal}) with Lyman-$\alpha$
in emission at $z=3.359$ has
been found serendipitously during a spectroscopic campaign with the
LBT MODS1 optical spectrograph on the COSMOS field (PI F. Fiore). 
After visual inspection of these new eight galaxies with HST F814W high
resolution imaging, none has been discarded due to
evident contamination by
foreground objects close to the line of sight of the $z\sim 3.3$
galaxies. The limitation of this procedure, however, is that only with
one HST band it is not possible to clearly spot possible blends of
multiple sources. We end up with 8 additional SFGs in the COSMOS field
which are enlarging the original sample adopted by \cite{grazian16}.

\subsection{The EGS field}

In order to improve the significance of our results by increasing the
number statistics, we included in the present analysis an additional
sample of star-forming galaxies in the Extended Groth Strip
(EGS/AEGIS) field (\cite{davis07}). An area of $\sim 600$
sq. arcmin. has been covered by two overlapping LBC pointings in the
UV band for a total of 6.3 hours (PI H. Yan). The UV depth is 29.6 at
a S/N=1, with an average seeing of 1.1 arcsec (\cite{grazLBC}).
The magnitude limit has been computed through a PSF-fitting method, as
described in the next section. Since
no observations in the R band have been acquired by LBC on this
specific pointing, we have used the R-band image taken by the ``Deep''
program of the CFHT Legacy Survey (hereafter CFHTLS\footnote{\em
http://www.cfht.hawaii.edu/Science/CFHTLS/}). In
particular, we have used the final release (T0007) of the CFHTLS,
which covers 1 sq. deg. of the EGS field at a depth of $r=25.6$ at an
80\% completeness level for point sources. The CFHTLS image has been
resampled to the same WCS astrometric grid of the LBC U-band image
with Swarp\footnote{\em http://www.astromatic.net/software},
preserving both the total fluxes of the objects and their original S/N
ratios. A large number of spectroscopic redshifts are available in the
EGS field from Steidel et al. (2003) and from the DEEP2 survey
(Cooper et al. 2006). An initial
sample of 16 galaxies has been selected with $3.27<z_{spec}<3.40$. One galaxy
(Westphal MD99) has been discarded since it has been detected at 24
micron by Rigopoulou et al. (2006), suggesting possible AGN activity,
and it is also contaminated by a foreground object. Thus, a final
sample of 15 galaxies has been extracted from the EGS database.

\subsection{The CANDELS/GOODS-North field}

In order to explore the LyC radiation of faint galaxies, we
successfully proposed an LBT Strategic Program (PI A. Grazian) during
the Italian LBT Call for Proposals 2012B, with the aim of obtaining
ultra deep imaging in the U and R bands of the CANDELS/GOODS-North
field (\cite{grogin11,koekemoer11,grazLBC}) with the LBC
instrument. The same area has been observed also by other LBT
partners (AZ, OSURC, and LBTO), for a total exposure time of 33 hours
in U band (seeing 1.1 arcsec) and 26 hours in R band (seeing 1.0
arcsec). The detailed description of this dataset is provided in a
different paper (\cite{grazLBC}) summarizing all the LBC deep
observations available in the CANDELS fields. The long
exposure time and the relatively good seeing obtained allow us to
reach a magnitude limit in the U band of 30.2 mag at S/N=1, resulting
in one of the deepest UV images acquired so far (\cite{ashcraft16}).
The LBC R band,
instead, reaches a magnitude limit of 27.8 at 80\% completeness level
for point sources.

The GOODS-North field is one of the premier targets of spectroscopic
campaigns with the Keck and other smaller telescopes, thanks to
the multiwavelength coverage of this area from X-ray to the Far IR. In
particular, \cite{barger08} collected 2907 spectroscopic redshifts for
galaxies and stars in the ACS footprint (\cite{giavalisco04}),
resulting in a 90\% spectroscopic completeness at $B_{435}\le 24.5$
mag. Additional information has been provided by
\cite{cooper11,pirzkal13,wirth15}, just to mention few examples of the
numerous work which provided spectroscopic redshifts in the
GOODS-North area. We have selected from this large database 11
galaxies with $3.27<z_{spec}<3.40$, the redshift range suitable for
LyC analysis with the LBC U-band filter. These galaxies are all the objects
in this tiny redshift interval, without any biases against or in favour of
Lyman-$\alpha$ emitters. One of these galaxies
(RA=12:37:33.2, DEC=+62:17:51.9) is extremely faint ($R\sim 28$) and
no meaningful constraint on the escape fraction can be achieved with
the present depth of the LBC U-band image. Moreover, this galaxy is
also contaminated by a brighter galaxy at less than 1.5 arcsec
distance, so we decided to remove this object from our $z\sim 3.3$
sample. In the end, the final sample in the GOODS-North field consists of 10
galaxies, of which two objects (with $ID>90000$ in Table
\ref{table:gal}) are outside the CANDELS area.

\subsection{The faint lensed galaxy sample}

To push our LyC analysis even deeper, we added the lensed galaxy by
\cite{amorin14}, at z=3.417 and magnified by a factor of 40$\pm 1$,
with intrinsic magnitude of $R_{delensed}=28.31$, corresponding to
$L=0.036L^*$ ($M_{1500}=-17.4$). The properties of this galaxy have
also been discussed by \cite{vanderwel13}.
Interestingly, this galaxy has been classified as a
metal poor galaxy by \cite{amorin14}, with $12+log(O/H)< 7.44$ (or
equivalently $Z<0.05Z_{\odot}$). In addition, it shows high ionization
conditions, with [OIII]/H$\beta$ ratio $>5$ and [OIII]/[OII]$>10$. We also
consider two lensed galaxies at $z>3$ with properties similar to this
object from the literature
(Christensen et al. 2012, Bayliss et al. 2014), where upper limits to the
LyC escape fraction have been provided. The properties of these galaxies
resemble the Green Peas selected by \cite{cardamone09}, and they are thus very
promising candidates to investigate for their possible contribution to
the \ion{H}{I} ionizing background, as recently proposed by
Izotov et al. (2016a), Vanzella et al. (2016),
Schaerer et al. (2016), Stark et al. (2017), Verhamme et al. (2017).

\subsection{The total sample}

In total, a sample of 79 star-forming galaxies within the redshift
range $3.27<z<3.4$ has been assembled, of which 45 have been already
presented in \cite{boutsia11} and \cite{grazian16}. The properties of
the additional 34 galaxies are summarized in Table \ref{table:gal}.
The results on the LyC emission properties are instead based on the
whole sample of these 79 star-forming galaxies with deep LBC data available
and are summarized in
the next sections. One of these 79 galaxies is the lens by \cite{amorin14}.
In the final sections, we also consider two additional lensed
galaxies at $z\sim 3$ from the literature, without LBC coverage
(Christensen et al. 2012, Bayliss et al. 2014).
They have been added to our sample
since they are relatively faint galaxies (magnified by strong lensing)
with an independent estimation of their LyC escape fraction.


\section{The Method}

\subsection{Estimating the Lyman Continuum Escape Fraction of $z\sim 3.3$
galaxies}

We adopt here the same methodology described by \cite{grazian16} in
order to estimate the individual values of the galaxy Lyman Continuum escape
fraction and the sample mean at $z\sim 3.3$. In particular, we
focus on the new sample of 34 galaxies with known spectroscopic
redshifts and with wide and deep imaging from COSMOS, EGS, and
CANDELS/GOODS-North. Briefly, we summarize here the adopted technique.

Following \cite{grazian16}, we started from images by LBC in the U and R
bands, which at $z\sim 3.3$ are sampling the rest frame wavelengths at
900 and 1500 {\AA}, respectively. The relevant galaxies have been
selected in a narrow redshift range ($3.27<z<3.40$), with the aim of
measuring, with the U-band filter of LBC, only the wavelength region
at $850\le \lambda\le 900$ {\AA} rest frame (see Fig. 1 of Boutsia et
al. 2011). The average Lyman continuum absorption by the IGM in the
spectra of $z\ge 3$ sources indeed increases rapidly towards bluer
wavelengths. For this reason the search for any detection of hydrogen
ionizing emission is quite inefficient below 850 {\AA} rest, which is
sampled by the LBC U-band filter in spectra of $z>3.4$ galaxies.
The red leakage of the U-band filter of LBC is
negligible\footnote{http://lbc.oa-roma.inaf.it/Filters/List.html}, and it
is not affecting the following results.

Crucially, almost all the new 34 galaxies have associated deep HST imaging from
the COSMOS, EGS, and CANDELS/GOODS-North surveys
(\cite{scoville,grogin11,koekemoer11,egs}) at several wavelengths
from the V to the H bands (for the CANDELS/GOODS-North field the B
band of HST is also available), which are essential in order to avoid
spurious contamination by foreground sources. In the latter field,
for example, there are three sources (ID=16479, 16169, 18979) with
bright galaxies with $z_{phot}\sim 0.7-2.2$ close to their line of
sights. Two (16169, 18979) are well separated ($\sim 1.5-1.8$
arcsec), while source ID=16479 is closer ($\sim 1.0$ arcsec) and
results in a blend in the LBC images. We will discuss the properties
of this source in the next sub-section.

The HST fields used here also have
deep X-ray observations by the Chandra satellite
(\cite{alexander03,nandra15,civano16}). We have used this information to avoid
possible contamination by faint AGNs. None of
our 34 galaxies shows any X-ray emission at the levels probed by the
Chandra images, thus excluding the presence of an X-ray strong AGN
nucleus inside.

As carried out in \cite{grazian16}, we have measured the ``relative''
escape fraction, which is defined as the fraction of emitted Lyman
continuum photons, IGM corrected, related to the observed fraction
of photons at 1500
{\AA} rest frame (\cite{steidel01,siana07}). The relative escape
fraction is usually derived from the observations of the flux ratio
between 900 and 1500 {\AA} rest:

\begin{equation}
f^{rel}_{esc}=\frac{(L_{1500}/L_{900})_{int}}{(F_R/F_U)_{obs}}exp(\tau_{900}^{IGM}) \,
\label{eqnfesc}
\end{equation}
\noindent
where $(L_{1500}/L_{900})_{int}$ is the ratio of the intrinsic
luminosities at 1500 and 900 {\AA} rest frame, $(F_R/F_U)_{obs}$ is
the ratio of the observed fluxes in the R and U band,
$\tau_{900}^{IGM}$ is the optical depth of the IGM at 900 {\AA} rest frame.
The absolute escape fraction $f^{abs}_{esc}$ is defined as
$\exp(-\tau_{\ion{H}{I}})\times 10^{-0.4 A_{900}}$, where the first factor
indicates the ISM extinction, while $A_{900}$ is the dust absorption coefficient
at 900 {\AA} rest frame.
In this paper, as in \cite{grazian16}, we adopt a value
$(L_{1500}/L_{900})_{int}$ of 3. In \cite{guaita16} we have explored the
dependencies of this quantity on the various population synthesis
models, concluding that a reasonable range for this ratio can be
between 2 and 7. In \cite{grazian16} we have considered the
implications of different assumptions of the intrinsic luminosity
ratio parameter on the estimations of the escape fraction and LyC
emissivity.

To correct for the mean effect of the IGM absorption we have adopted the
recent estimates by \cite{worseck14a} and \cite{inoue14}. Folding
the IGM absorption at $z=3.3$ with the U-band filter of LBC and
adopting the same procedure outlined in \cite{boutsia11}, we obtain a
mean IGM transmission $<exp(-\tau_{900}^{IGM})>=0.28$ at $z\sim 3.3$.
In the next section we explore the implications of the
variance of the IGM absorption along multiple lines of sight, using
the same methodology adopted in \cite{grazian16}.

For each galaxy we have computed the observed flux ratio
$(F_R/F_U)_{obs}$ from the LBC images in the U and R bands in the same
way as in \cite{grazian16}. Briefly, we detect the $z\sim 3.3$ galaxy
in the R band using SExtractor (\cite{sex}), and measure the flux in
the U band assuming the same profile of the galaxy in the R band
through the ConvPhot software (\cite{desantis07}), taking into account
the transfer function (convolution kernel) to change the spatial
resolution of the R band into the U-band one. As discussed in
\cite{grazian16}, this PSF-matched photometry allows us to derive
errors on fluxes which are typically a factor of 1.7 smaller than the
ones measured by SExtractor through aperture photometry. In case of
non-detection in the U band, a 1-$\sigma$ upper limit to the flux in
this filter is set by the flux error provided by ConvPhot. Assuming an
average IGM transmission of 0.28 at $z\sim 3.3$, we compute the relative escape
fraction for each individual galaxy adopting Eq.\ref{eqnfesc}, and
provide the relevant quantities in Table \ref{table:gal}.

\begin{table*}
\caption{The new galaxies in the COSMOS, EGS, and CANDELS/GOODS-North
fields used to derive the LyC relative escape fraction}
\label{table:gal}
\centering
\begin{tabular}{r | c c c c c c c c c}
\hline
\hline
$ID$ & $Field$ & $RAD$ & $Dec$ & $z_{spec}$ & $Rmag$ & $Umag$ & $M_{1500}$
 & $f^{rel}_{esc}$ & $Cont$ \\
   &     & (deg)    & (deg)      &     &       (AB) & (AB) & &      &           
\\
\hline
 53167 & COSMOS & 149.852446 & +2.397032 & 3.359 & 22.95 & $\ge$ 29.38 & -22.75 & $\le$ 0.029 & No \\
218783 & COSMOS & 149.920820 & +2.387060 & 3.297 & 24.49 & $\ge$ 29.39 & -21.21 & $\le$ 0.118 & No \\
223954 & COSMOS & 149.831880 & +2.404150 & 3.371 & 24.77 & $\ge$ 29.58 & -20.93 & $\le$ 0.128 & No \\
219315 & COSMOS & 149.839520 & +2.388460 & 3.363 & 25.21 & $\ge$ 29.70 & -20.49 & $\le$ 0.171 & No \\
215511 & COSMOS & 149.848260 & +2.376170 & 3.363 & 25.24 & $\ge$ 29.70 & -20.46 & $\le$ 0.176 & No \\
220771 & COSMOS & 149.836280 & +2.393190 & 3.359 & 24.84 & $\ge$ 29.22 & -20.86 & $\le$ 0.188 & No \\
217597 & COSMOS & 149.865340 & +2.382770 & 3.284 & 25.73 & $\ge$ 29.88 & -19.97 & $\le$ 0.233 & No \\
211934 & COSMOS & 149.847250 & +2.364040 & 3.355 & 26.86 & $\ge$ 29.64 & -18.84 & $\le$ 0.827 & No \\
\hline
59330$^a$  & COSMOS & 150.077045 & +2.360797 & 3.417 & 24.15 & $\ge$ 28.93 & -17.40 & $\le$ 0.232 & No \\
\hline
6328  & EGS & 214.586647 & +52.512559 & 3.357 & 24.26 & $\ge$ 29.89 & -21.44 & $\le$ 0.060 & No \\
9524  & EGS & 214.500676 & +52.477565 & 3.295 & 24.06 & $\ge$ 29.54 & -21.64 & $\le$ 0.069 & No \\
6439  & EGS & 214.552783 & +52.511284 & 3.392 & 24.15 & $\ge$ 29.56 & -21.55 & $\le$ 0.073 & No \\
724   & EGS & 214.501128 & +52.570535 & 3.353 & 24.76 & $\ge$ 30.06 & -20.94 & $\le$ 0.081 & No \\
6922  & EGS & 214.461916 & +52.505537 & 3.351 & 24.71 & $\ge$ 29.95 & -20.99 & $\le$ 0.086 & No \\
4580  & EGS & 214.508258 & +52.531509 & 3.279 & 24.79 & $\ge$ 29.90 & -20.91 & $\le$ 0.096 & No \\
8642  & EGS & 214.460301 & +52.487802 & 3.285 & 24.50 & $\ge$ 29.60 & -21.20 & $\le$ 0.098 & No \\
5197  & EGS & 214.283603 & +52.524388 & 3.271 & 24.52 & $\ge$ 29.15 & -21.18 & $\le$ 0.150 & No \\
6612  & EGS & 214.500033 & +52.509127 & 3.340 & 25.89 & $\ge$ 30.45 & -19.81 & $\le$ 0.162 & No \\
2783  & EGS & 214.554012 & +52.549350 & 3.272 & 25.70 & $\ge$ 30.02 & -20.00 & $\le$ 0.200 & No \\
12335 & EGS & 214.455464 & +52.446348 & 3.287 & 25.32 & $\ge$ 29.61 & -20.38 & $\le$ 0.205 & No \\
13346 & EGS & 214.413697 & +52.433529 & 3.307 & 24.73 & $\ge$ 28.87 & -20.97 & $\le$ 0.237 & No \\
7690  & EGS & 214.249912 & +52.497386 & 3.272 & 25.18 & $\ge$ 28.77 & -20.52 & $\le$ 0.392 & No \\
11372 & EGS & 214.407226 & +52.458201 & 3.294 & 25.36 &       28.48 & -20.34 &       0.603 & No \\
14043 & EGS & 214.376422 & +52.424380 & 3.344 & 25.18 & $\ge$ 28.29 & -20.52 & $\le$ 0.610 & No \\
\hline
16428 & GDN & 189.216181 & +62.254036 & 3.333 & 23.91 & $\ge$ 29.82 & -21.79 & $\le$ 0.048 & No \\
19977 & GDN & 189.179397 & +62.276656 & 3.363 & 24.38 & $\ge$ 30.03 & -21.32 & $\le$ 0.061 & No \\
12305 & GDN & 189.219779 & +62.227532 & 3.369 & 25.09 & $\ge$ 30.35 & -20.61 & $\le$ 0.087 & No \\
4341  & GDN & 189.290579 & +62.171244 & 3.384 & 24.68 & $\ge$ 29.81 & -21.02 & $\le$ 0.094 & No \\
14132 & GDN & 189.206264 & +62.239252 & 3.367 & 25.44 & $\ge$ 30.23 & -20.26 & $\le$ 0.135 & No \\
91200 & GDN & 189.311200 & +62.311020 & 3.390 & 25.76 & $\ge$ 30.21 & -19.94 & $\le$ 0.184 & No \\
16169 & GDN & 189.215365 & +62.252800 & 3.364 & 25.42 & $\ge$ 29.67 & -20.28 & $\le$ 0.214 & No \\
18979 & GDN & 189.296340 & +62.270650 & 3.350 & 25.80 & $\ge$ 29.90 & -19.90 & $\le$ 0.246 & No \\
91000 & GDN & 188.999330 & +62.202417 & 3.300 & 26.53 & $\ge$ 30.21 & -19.17 & $\le$ 0.377 & No \\
16479 & GDN & 189.194532 & +62.254757 & 3.371 & 25.00 &       27.79 & -20.70 &       0.821 & Yes \\
\hline
\hline
\end{tabular}
\\
The galaxies have been sorted in ascending order of relative escape
fraction in each field (COSMOS, EGS, GDN=CANDELS/GOODS-North). The
identification number, $ID$, is the LBC one for COSMOS and EGS fields,
while it is the CANDELS official identifier for the GOODS-North field.
Two sources, indicated by $ID>90000$, are outside the
CANDELS/GOODS-North area. The non detections (i.e. $S/N<1$) in the LBC
U-band, for which an upper limit (at $1\sigma$) to the
relative escape fraction has
been provided, are indicated by the $Umag\ge$ and $f^{rel}_{esc}\le$
symbols, respectively. The last column, $Cont$, indicates whether a
galaxy is possibly contaminated by foreground objects. Notes on
individual objects: (a): object ID=59330 in COSMOS is the lensed
galaxy studied by \cite{vanderwel13} and by \cite{amorin14}. The U-
and R-band magnitudes are the observed photometry for ID=59330,
while the absolute
magnitude provides the intrinsic luminosity after the correction by a
magnification factor of $40\pm 1$.
\end{table*}

\subsection{Detailed analysis of the possible LyC emitters}

In \cite{grazian16} we have discussed the issue of foreground
contaminants mimicking significant LyC detection in the COSMOS
field. In Table \ref{table:gal} there are two galaxies with
significant ($\sim 60-80\%$) LyC relative escape fraction, ID=11372 in
the EGS field and ID=16479 in GOODS-North. In order to check their
reliability, we carried out some checks, as done in \cite{grazian16}. In
particular, we used HST data from the CANDELS and AEGIS surveys in order to
evaluate the reliability of these LyC emitter candidates.

\begin{figure*}
\includegraphics[width=18cm,angle=0]{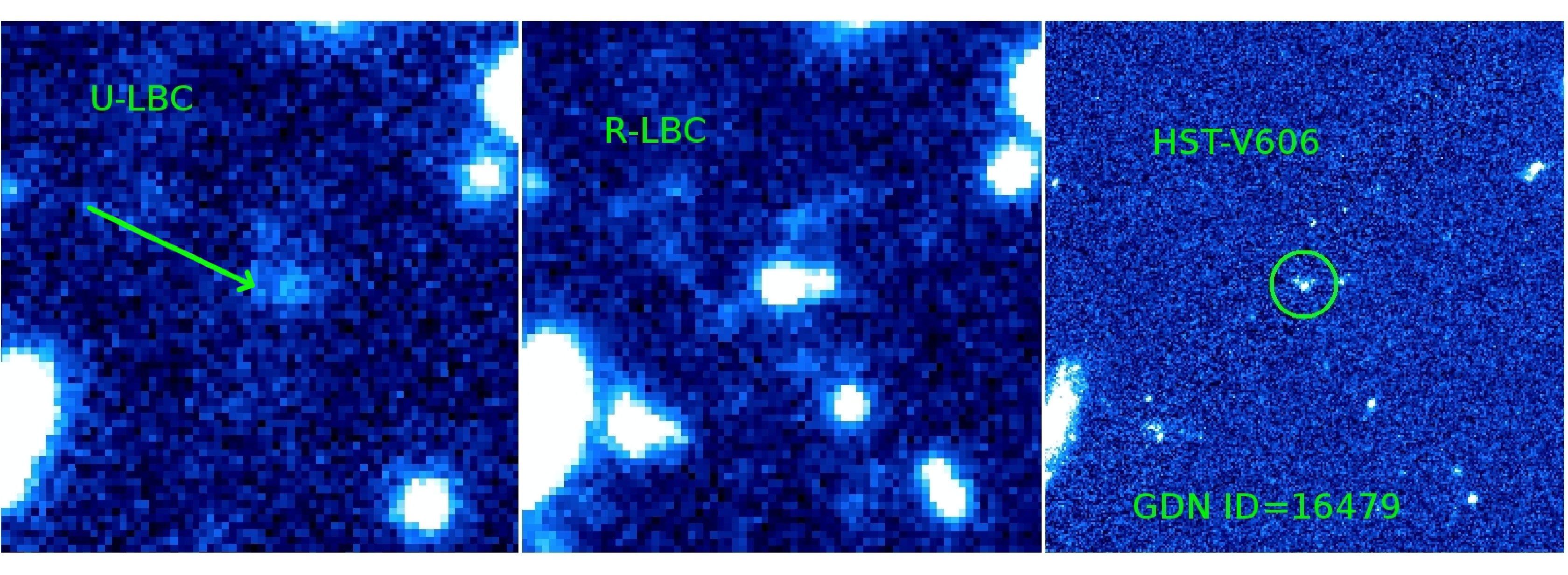}
\caption{
The cutouts of source ID=16479 of the CANDELS/GOODS-North field in the
U band (left) and in the R band (center) by the LBC instrument. The
cutout on the right shows the same sky area observed by HST in the
V606 filter. The position of the galaxy with spectroscopic redshift
at $z=3.371$ is indicated by the green circle. The green arrow in the
U band marks the position of the centroid of source ID=16479
derived from the R-band image.
At $\sim 1.0$ arcsec from the
$z\sim 3.3$ source, towards the right, there is a faint galaxy at
$z_{phot}=2.2$ which
is clearly detected in all the three bands. Close to the main target
there are also two very faint blobs in the HST image, which are
blended with the $z\sim 3.3$ galaxy in the ground-based images by LBC.
The size of each image is 17 $arcsec$ by 17 $arcsec$.
}
\label{obj16479ima}
\end{figure*}

\begin{figure*}
\includegraphics[width=18cm,angle=0]{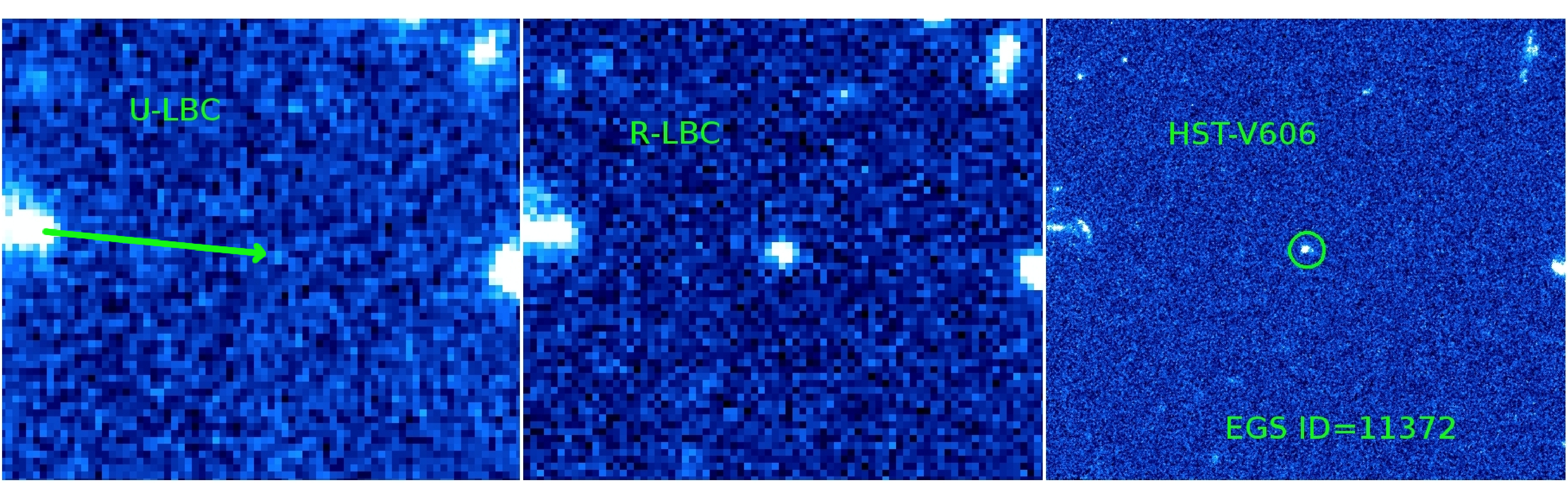}
\caption{
The cutouts of source ID=11372 of the EGS field in the U band (left)
and in the R band (center) by the LBC instrument. The cutout on the
right shows the same sky area observed by HST in the V606 filter. The
position of the galaxy with spectroscopic redshift at $z=3.294$ is
indicated by the green circle.
The green arrow in the
U band marks the position of the centroid of source ID=11372
derived from the R-band image.
In the LBC U band there is a marginal
(1.5 $\sigma$) detection. The size of each image is 17 $arcsec$ by 17
$arcsec$.
}
\label{obj11372ima}
\end{figure*}

Figure \ref{obj16479ima} shows the galaxy ID=16479 in the
CANDELS/GOODS-North field in the U and R band by LBC and in the HST
V606 filter (the closest band to the LBC-R one). At the right of the
$z=3.371$ galaxy (identified by a green circle in the HST thumbnail)
there is a faint source at $\sim 1.0$ arcsec distance, which is clearly
detected in the three bands. If we assume that the two sources are at
the same redshift ($z=3.371$) and compute their escape fraction, we find
$f^{rel}_{esc}=206\%$. If we evaluate the escape fraction only of the
fainter galaxy to the right (outside the green circle) and
assuming it is at $z\sim 3.3$, we end up with $f^{rel}_{esc}=627\%$. In
this case, similar to that already reported by \cite{vanzella12} and
\cite{grazian16}, the so-called ``local'' escape fraction (i.e. calculated
in a small region showing LyC emission) is not
compatible with the stellar population of a $z\sim 3.3$ galaxy,
indicating a clear sign of
contamination by a lower-z interloper. In fact, even assuming a higher
ionizing efficiency (i.e. a lower $(L_{1500}/L_{900})_{int}$ value of
2.0, instead of 3.0) and a more transparent line of sight with
$exp(-\tau_{IGM})=0.68$ (i.e. the upper value of the $z\sim 3.3$ IGM
transmission at 95\% c.l.), the resulting ``local'' escape fraction is
170\%, confirming the interloper interpretation for this object.
Indeed, the brighter galaxy at the center of the green circle
in Fig. \ref{obj16479ima} has different colors with respect to the
faint blob on the right, which is thus probably at a different
redshift. This system has already been studied by \cite{reddy06},
labeled as BX1334. Interestingly, it has been selected as a BX galaxy
(with colors compatible with galaxies at photometric redshifts between
2.0 and 2.7, see \cite{adelberger04}), and it has NIR colors which are
compatible with the distant red galaxy (DRG) criterion ($R-K\ge 1.3$ mag,
$z\sim 2$ according to \cite{drg}).
From the CANDELS photometry a
redshift of $z_{phot}=2.2$ has been derived for this close-by source,
indicating that this is plausibly a lower-z contaminant.

Using HST data in the V606 band,
it is possible to compute the relative escape fraction of the $z\sim
3.3$ galaxy reducing the contamination by the foreground interloper on
the right. Adopting as a prior for ConvPhot the HST image we obtain
an U-band magnitude of 27.79, corresponding to
$f^{rel}_{esc}=82\%$. This could thus be a genuine LyC emitter at $z\sim
3.3$. However, close to the $z\sim 3.3$ galaxy there are also two very
faint blobs in the HST image, which are blended with the main target
in the ground-based images by LBC. A faint off-center emission in the
U band of LBC has been detected, and this is possibly associated to
the faint blob close to the $z=3.371$ galaxy (towards the
right). There is thus an indication of a possible contamination by
foreground objects, even if in this case no firm conclusions can be
drawn.

The second object ID=11372, shown in Fig. \ref{obj11372ima}, is at best
marginally detected at 1.5 $\sigma$ in the U band, and its relative
escape fraction is $f^{rel}_{esc}\sim 60.3\%$. This galaxy has
been originally studied by \cite{steidel03} as the object C029 of the
Westphal field. Its morphology in the HST V606 band is almost
point-like, but no X-ray detection has been found for this galaxy, and
thus an AGN classification for this object cannot be confirmed. Even for
this object there is a hint in the HST image of a very faint blob
close to the $z\sim 3.3$ galaxy (at $\sim$0.1 arcsec towards the
right), which can act as a possible faint contaminant, compatible with
the marginal (1.5 sigma) detection in the U band. However, no firm
conclusions can be drawn for this galaxy, and deeper data with LBC in
the U band and HST in the V606 band are needed in order to understand
its nature. It is possible indeed that the distribution of the escape
fraction of $z>3$ galaxies is bimodal, with the bulk of the star-forming
population showing small values of LyC emission, while few objects
emit large amount of ionizing radiation (e.g. Ion2, Vanzella et al. 2016).

In Table \ref{table:gal}, we mark the ID=16479 galaxy as possibly
contaminated by a foreground object, while the galaxy ID=11372, where
this hypothesis is weak, is considered as a possible LyC emitter,
pending deeper observations. In the following sections we investigate
the properties of the whole star-forming galaxy population at $z\sim
3.3$, relying on the stacking procedure of the LBC images.

In summary, the properties of the new sample of 34 galaxies at
$z\sim 3.3$ in the
COSMOS, EGS, and CANDELS/GOODS-North fields have been collected in
Table \ref{table:gal}. They have been merged with the 45 SFGs at the
same redshifts studied in \cite{grazian16} from the VUDS/LBC-COSMOS
database, resulting in a total sample of 79 galaxies. However, not all
of them have been used to compute the \ion{H}{I} ionizing emissivity of the
SFG population at $z\sim 3$. We have indeed discarded 9 galaxies
possibly affected by contamination due to foreground objects. Of this
sub-sample, eight galaxies were discussed in detail in
\cite{grazian16} and an additional galaxy has been discarded from
Table \ref{table:gal}, i.e. ID=16479 in the CANDELS/GOODS-North field.
The galaxy ID=59330 in the COSMOS field is the lensed object studied by
\cite{amorin14} and it will not be considered in the stacking procedure,
which is restricted to the 69 galaxies of luminosity $L\gtrsim 0.2L^*$.


\section{Results}

We consider here the stack of the cleaned sample in
the U and R bands by LBC and provide the mean $f^{rel}_{esc}$ values
for the whole population of 69 galaxies not affected by contamination
and for various sub-samples at different R-band magnitudes.

\begin{figure*}
\centering
\includegraphics[width=14cm,angle=-90]{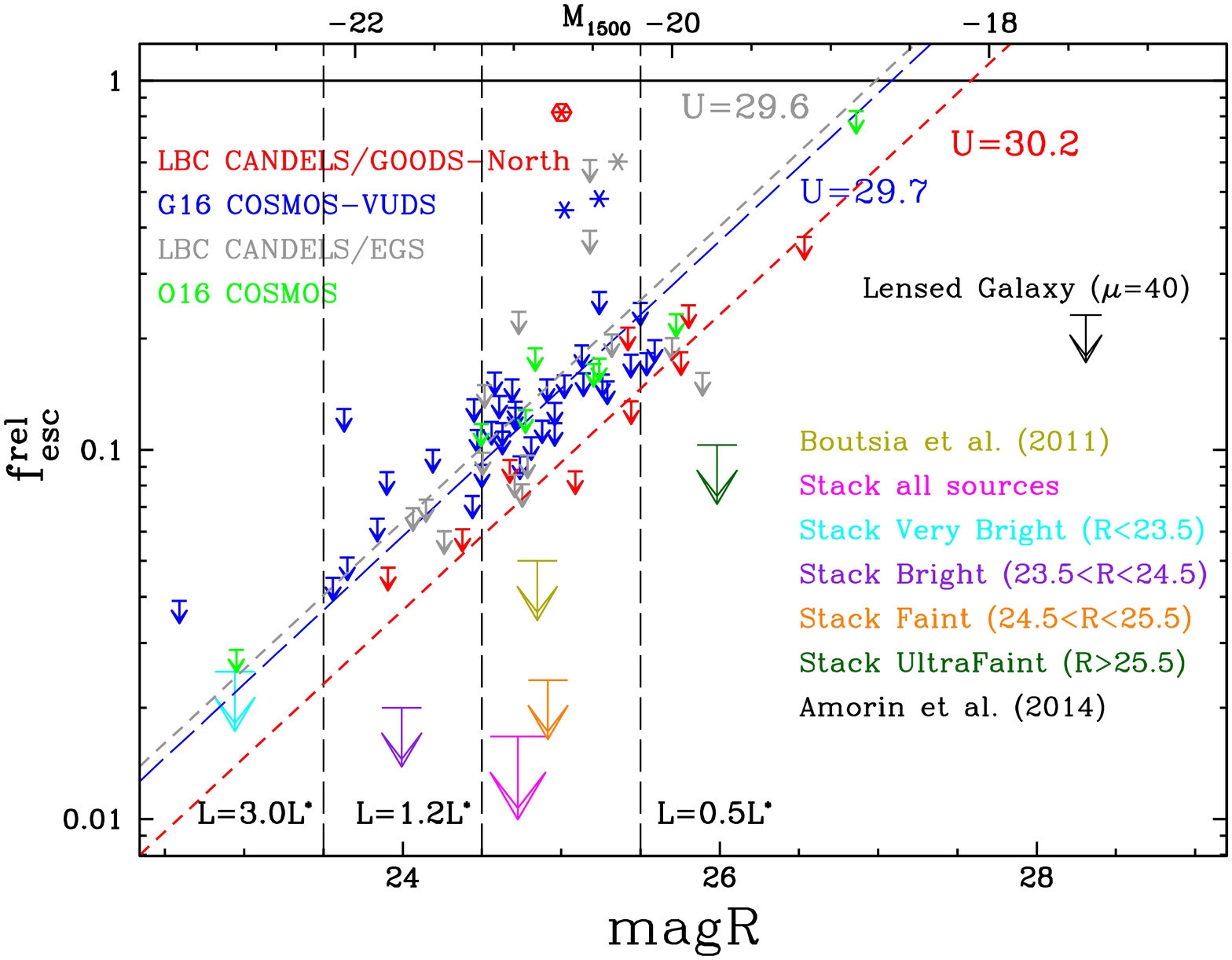}
\caption{
The measured values and the upper limits at 1 $\sigma$ for the LyC
relative escape fraction of galaxies at $z\sim 3.3$ in the COSMOS
(blue, green), EGS (grey), and CANDELS/GOODS-North fields (red). Small
descending arrows show the upper limits of $f^{rel}_{esc}$ associated
to individual galaxies. Asterisks show
the galaxies with detection in the U band and a LyC emission above
1$\sigma$. The red hexagon indicates a likely foreground contamination
in the CANDELS/GOODS-North field, associated to galaxy ID=16479. The lensed
galaxy by \cite{amorin14}, magnified by a factor $\mu=40\pm 1$, is
represented by a black arrow at its intrinsic (de-magnified)
magnitude $R=28.31$. The grey, blue, and red dashed lines
show the dependency of the escape fraction on the R-band magnitude
derived by adopting the
$1\sigma$ depth of the U band in the EGS, COSMOS, and CANDELS/GOODS-North
fields, respectively.
The large arrows (cyan, purple, magenta,
orange, gold, dark green) represent the limits to $f^{rel}_{esc}$
derived by stacking the $z\sim 3.3$ galaxies of \cite{grazian16} and
of this paper in different R-band magnitude intervals.
}
\label{magfesc}
\end{figure*}

Fig. \ref{magfesc} summarizes the $f^{rel}_{esc}$ properties of the
cleaned sample as a function of the observed R-band magnitude. In this
plot, the blue symbols indicate the VUDS/LBC-COSMOS galaxies published in
\cite{grazian16}, while the green, red, and grey arrows are for
sources in the COSMOS (Onodera et al. 2016), CANDELS/GOODS-North, and EGS
fields, respectively. The small downward arrows indicate 1$\sigma$ upper limit
for the individual $f^{rel}_{esc}$, while asterisks show detections in
LyC above 1$\sigma$. The hexagon upon the red asterisk indicates the
galaxy ID=16479 in the CANDELS/GOODS-North field, which is probably
affected by foreground contamination. The grey, blue, and red dashed lines
show the escape fraction vs R-band magnitude corresponding to the
$1\sigma$ depth of the U band in the EGS, COSMOS, and CANDELS/GOODS-North
fields, respectively (U=29.6, 29.7, and 30.2), derived using
Equation \ref{eqnfesc}. The scatter of the observed galaxies around these
lines is due to the small differences in depth on the images and on the
different morphologies of the sources (more extended galaxies have
slightly shallower magnitude limits).

\subsection{Constraints on the escape fraction from stacking}

We then proceed with the stacking of the LBC U and R bands of the
clean sample in different luminosity intervals. We repeat the same
procedures outlined in \cite{grazian16}, which we summarize in the
following. We cut a small thumbnail ($\sim 45{``}\times 45{``}$)
of the LBC U- and R-band images centered on each $z\sim 3.3$ galaxy,
we mask all the nearby sources, detected in the R band, leaving
unmasked the central region, then we compute the weighted mean of all
the thumbnails in a given band using the inverse variance map as a
weight. We measure the $f^{rel}_{esc}$ on the stacked images adopting
the same technique outlined above, i.e. using ConvPhot.

We first proceed with the stacking in different R-band magnitude
intervals, then we carry out the sum of all the galaxies in
\cite{grazian16} and in Table \ref{table:gal}, except for the
contaminated object (ID=16479) in the CANDELS/GOODS-North field and
the lensed galaxy by \cite{amorin14}. We do not achieve any
significant detection in the U-band above $1\sigma$ in the stacked
images, resulting in several upper limits to the LyC escape
fraction. We plot the results of the stacking in different luminosity
intervals in Fig. \ref{magfesc} with large arrows, and summarize all
the relevant values in Table \ref{table:fesc}.

\begin{figure}
\includegraphics[width=9cm,angle=0]{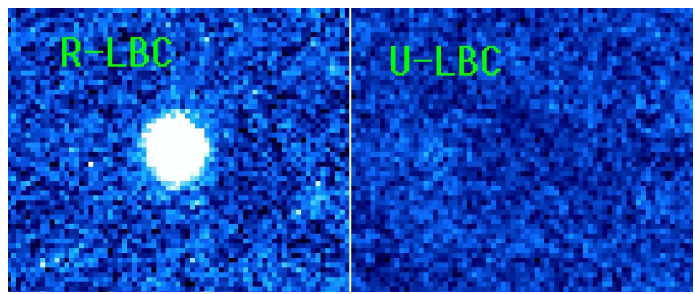}
\caption{
Stack of the 69 galaxies with $3.27\le z_{spec}\le 3.40$ in the R
({\em left}) and U-bands ({\em right}) for the COSMOS, EGS, and
CANDELS/GOODS-North fields. The size of each image is 16 arcsec
by 14 arcsec.
}
\label{stackall}
\end{figure}

We do not detect for the stacks significant dependencies of
$f^{rel}_{esc}$ on luminosity at $R\le 25.5$, corresponding to $L>0.5
L^*$, where a constant value of 2-2.5\% has been determined (cyan,
purple, and orange arrows in Fig. \ref{magfesc}). Stacking of all the
69 galaxies (we do not consider the lensed galaxy by
Amorin et al. 2014) in U and R band resulted in a non detection in the U
band to a level of $U=31.74$ at 1$\sigma$ and a clear detection in the
R band of $R=24.73$ at S/N=414, as shown in Fig. \ref{stackall}.
This translates into a limit to the
observed ratio between R and U band of $F_R/F_U\ge 640.2$,
which is equivalent to an upper limit to
the relative escape fraction of $f^{rel}_{esc}\le 1.7\%$ at 1$\sigma$
level at z=3.3 (magenta
arrow in Fig. \ref{magfesc}). This is the most stringent limit to the relative
escape fraction for $z>3$ star-forming galaxies. The inclusion of the
three galaxies detected in the U band (two galaxies discussed in
\cite{grazian16} and ID=16479 in the CANDELS/GOODS-North field) does
not change significantly the results of the stacking ($f^{rel}_{esc}\le 1.8\%$).

The uncertainties on the relative escape fraction of the stack due to
the IGM variance has been derived following the same procedures
outlined in \cite{grazian16}. Considering 69 lines of sight, the mean
IGM transmission, folded with the U-band filter by LBC, has an
uncertainty of $\sigma=0.023$. Using this error, the 68\% confidence
level for our value of $f^{rel}_{esc}$ is between 1.54 and 1.82\%.
In the following, we will use the derived limit $f^{rel}_{esc}\le 1.7\%$,
without considering its small associated uncertainty.

\subsection{Constraints on the escape fraction from strong
gravitational lensing}

Ultra-deep UV imaging in blank fields with LBC are reaching a magnitude
limit of $\sim 30$ (AB), but these depths do not allow us to push the
constraints of the escape fraction below $\sim 10\%$ at
$M_{1500}\gtrsim -20$. On the other hand, relatively deep LBC data in the U
and R bands have been used to put a limit to the LyC escape fraction
of the lensed galaxy by \cite{amorin14}, thus allowing us to explore
the regime of sub-$L^*$ galaxies at $z>3$. A 1$\sigma$ upper limit to
the escape fraction of 23.2\% has been derived at $M_{1500}=-17.4$,
which can add new observational constraints to the theoretical models
predicting that sub-$L^*$ galaxies at high-z have high escape
fractions, and are thus the main drivers of the reionization of the
Universe (e.g. \cite{ferrara13,ellis14,bouwens15,finkelstein15}).

Other examples of strongly magnified galaxies at $z\gtrsim 3$ can be
found in the literature. \cite{christensen12} reported the case of a
lensed LBG (SMACS J2031.8-4036 ID 1.1) at $z\sim 3.5$ with a LyC escape
fraction estimate from their X-shooter spectrum. This galaxy has
an absolute magnitude, after correction for magnification, of
$M_{1500}=-20.4$ and the constraint on the escape fraction
is $f^{rel}_{esc}\le 3-11\%$. Their conservative upper
limit for the escape fraction, thanks to strong lensing magnification,
is fully consistent with the individual limits on $z\sim 3.3$ galaxies
at the depth of the CANDELS/GOODS-North U-band observations of this
paper. The spectral properties of this peculiar galaxy are similar to
the ones observed in Ion2 by \cite{vanzella16}, and usually associated with
possible LyC emissivity: this galaxy indeed shows low metallicity
($Z=0.1 Z_\odot$) and a high ionization parameter ($logU=-2.1$),
presenting high [OIII]/[OII] and [OIII]/H$\beta$ line ratios. Following
recent claims (\cite{vanzella16,izotov16,schaerer16,nakajimaouchi,
nakajima16}), a much larger escape fraction should be expected
for this object. Due to its bright observed magnitude ($V_{606}=21.19
\pm 0.10$) this galaxy could be an ideal target for future deep
follow-up spectroscopy with efficient UV spectrographs in the LyC region.

Another interesting object is the Green Pea SGASJ105039.6+001730 at
z=3.6252 studied by Bayliss et al. (2014). This galaxy has a
magnification factor of $\mu \ge 30$, with an absolute magnitude of
$M_{1500}=-20.5$ after corrections for strong lensing. Like the previous
example, this object also shows 
line ratios ([OIII]/[OII]$>10$ and [OIII]/H$\beta>10$)
and is expected to emit ionizing radiation with a large
escape fraction, due to the significant ionization
efficiency. Surprisingly, the spectroscopic data indicate that the
neutral column density
is $\log\ N_{\ion{H}{I}}>21.5$ and thus the escape fraction is negligible
($f^{rel}_{esc}<1$\%). It is
worth noting that the profile of the Lyman-$\alpha$ line in
emission is also compatible with an associated absorption, in
agreement with the measured high column density of neutral hydrogen.
If its low $f^{rel}_{esc}$ is confirmed, this result gives stringent
limits to the LyC escape fraction at $z\ge 3$ which are comparable
with the constraints we have obtained in this work by stacking 69
galaxies ($\le 1.7\%$). It thus stresses the importance of the lensing
magnification for this type of studies, as also shown by
\cite{vasei16} at $z\sim 2$. Similar studies will be important in the
future to extend the investigated range of $f^{rel}_{esc}$ at very low
luminosities.

\begin{table}
\caption{The escape fraction of stacks}
\label{table:fesc}
\centering
\begin{tabular}{c c c c c}
\hline
\hline
Sample & $N_{gal}$ & Rmag & $M_{1500}$ & $f^{rel}_{esc}$ \\
       &          & (AB) & & (1$\sigma$) \\
\hline
Very bright ($R\le 23.5$) & 2 & 22.94 & -22.8 & $\le$0.025 \\
Bright ($23.5<R\le 24.5$) & 16 & 23.99 & -21.7 & $\le$0.020 \\
Faint ($24.5<R\le 25.5$) & 42 & 24.92 & -20.8 & $\le$0.024 \\
Ultra-faint ($R>25.5$) & 9 & 25.98 & -19.7 & $\le$0.103 \\
\hline
Stack all sources & 69 & 24.73 & -21.0 & $\le$0.017 \\
\hline
Lensed ($R_{delensed}=28.31$) & 1 & 28.31 & -17.4 & $\le$0.232 \\
\hline
Christensen et al. (2012) & 1 &  & -20.4 & $\le$0.11 \\
Bayliss et al. (2014) & 1 &  & -20.5 & $\le$0.01 \\
\hline
\end{tabular}
\\
The lensed galaxy with $R_{delensed}=28.31$ is described in \cite{amorin14}.
The galaxy by Bayliss et al. (2014) has
$logN(\ion{H}{I})\ge 21.5$, resulting in a null escape fraction.
We provide here
a very conservative upper limit of 1\% for this object.
\end{table}

\subsection{Does the escape fraction depend on luminosity ?}

In \cite{grazian16} we have shown that star-forming galaxies brighter
than 0.5 $L^*$ at $z\sim 3$ cannot produce the UV background (UVB) as measured
by the Lyman forest. Here we explore the role of fainter galaxies.

In Fig. \ref{magfesc} we observe an apparent trend of looser limits of
the escape fraction with fainter R-band magnitudes. This correlation
is clearly due to the U-band depth of our imaging database. At a given
luminosity, indeed, the measured upper limits on the escape fraction
of each individual galaxy depend on the different depths of the
COSMOS, EGS, and CANDELS/GOODS-North fields, and they are not related
to an intrinsic property in the population of star-forming galaxies.

The stacking at $R\le 25.5$ indicates that $f^{rel}_{esc}$ is below
$\sim$2-2.5\% level, while at fainter magnitudes the constraints
are less well-determined, with $f^{rel}_{esc}\le 10\%$ at $25.5\le R\le 27.5$
and $f^{rel}_{esc}\le 23\%$ at $R\sim 28.3$ mag, even if in the latter
case only one galaxy has been used to provide this limit, and
uncertainties due to the IGM variance still hold.

Leethochawalit et al. (2016), using the gas covering fraction
($f_{cov}$) of low ionization gas as an indirect proxy for the galaxy
escape fraction ($f^{abs}_{esc}\sim 1-f_{cov}$ neglecting dust
absorption), found that there is a weak anti-correlation of the
supposed $f^{rel}_{esc}$ with the SFR, with a typical value of
$f^{rel}_{esc}\sim 20\%$ for SFR of the order of 10-100 $M_\odot/yr$
at $4<z<5$. The range in SFR of their galaxies translates into an
R-band magnitude of $R\sim 25-28$, so it is interesting to note that
the value of $f^{rel}_{esc}$ inferred in their work is slightly
higher than what we measure.

A rough explanation is that the gas covering fraction can provide only an
upper limit to the LyC escape fraction, which in fact could be
significantly lower. A notable example is the lensed galaxy by
\cite{vasei16}: it has a low gas covering fraction, mimicking an
escape fraction of $f^{rel}_{esc}\sim 40\%$, but when a direct
measurement is carried out, a $3\sigma$ upper limit of 8\% has been
derived, significantly lower than the value inferred from the gas
covering factor (see also Dijkstra et al. 2016). These comparisons point
thus to a possible inconsistency between the indirect results derived
from the low covering fraction and the direct measurements of the LyC
escape fraction. It is thus desirable that future claims of
significant LyC emission be substantiated by direct detections.
However, it is useful to recall here that the limits by
\cite{amorin14} or \cite{vasei16} are based on single non-detections,
which could be due also to a line of sight with substantial IGM
attenuation, or to a patchy ISM. Larger databases of lensed,
intrinsically faint galaxies are needed to distinguish these
alternative hypotheses.

\begin{figure}
\includegraphics[width=7cm,angle=-90]{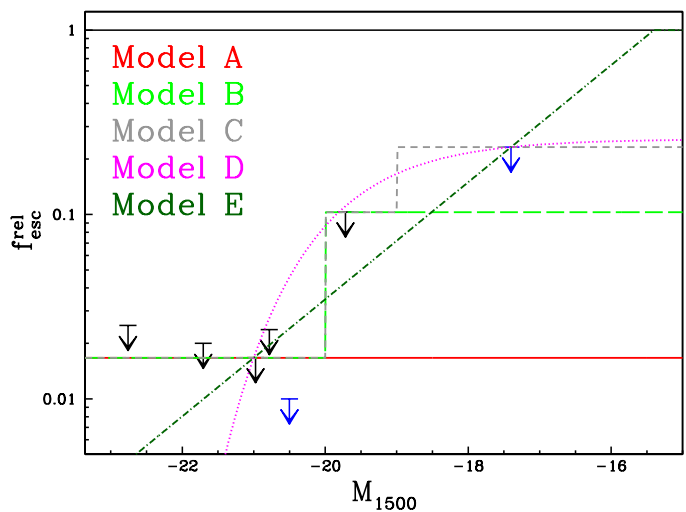}
\caption{
The dependence of the relative escape fraction from the absolute
magnitude in the UV $M_{1500}$. The black arrows are the upper limits
to the relative escape fractions derived from the stacks of LBC data in
this paper, summarized in Table \ref{table:fesc}. The observed trend
of looser upper limits with fainter magnitudes is due to the
limited depth of the data in the UV wavelengths, and plausibly it is
not related to a physical property of the population of star-forming
galaxies. The blue arrows summarized the values available in
literature for individual lensed galaxies, by \cite{amorin14} and
\cite{bayliss14}, in the luminosity ranges not covered by our data.
For the \cite{bayliss14} source, the limit of
1\% has been plotted only for graphical reasons, but could be
significantly lower. The different curves show analytic functions
which have been adopted to infer the evolution of the relative escape
fraction from luminosity. These functions have been used in order to
check for the contribution of galaxies at different luminosities to
the observed UVB.
}
\label{evolfesc}
\end{figure}

The data in Table \ref{table:fesc} have been summarized in
Fig. \ref{evolfesc}. Black arrows are the limits derived in this paper
with LBC data through stacking, while blue arrows report constraints
of individual lensed galaxies from the literature (Christensen et al. 2012,
Amorin et al. 2014, Bayliss et al. 2014). We explore here different
dependencies of the relative escape fraction with luminosity, with the
aim of computing the total ionizing emissivity of faint galaxies. The
galaxy by Bayliss et al. (2014) has $\log\ N(\ion{H}{I})\ge 21.5$,
resulting in a
very small ($f^{rel}_{esc}<1\%$) escape fraction. We provide here a very
conservative upper limit of 1\% for this galaxy in Fig. \ref{evolfesc}
only for graphical reasons. The real
emissivity value will be probably lower than this limit.

Different parameterizations of the escape fraction as a function of
luminosity have been explored in Fig. \ref{evolfesc}. It is worth
stressing here that these dependencies are not the best fit to the
observed data, but they are simply analytic functions (constant,
exponential or power law) which are
reasonably consistent with the observed stack limits, i.e. with looser
constraints at fainter luminosities. These functions are
not physically motivated, but they explore possible dependencies of
the escape fraction on luminosity $f^{rel}_{esc}(L)$ allowed by the
observed constraints. Indeed, it is useful to recall that the
contribution by bright galaxies does not change significantly if we
impose $f^{rel}_{esc}<1\%$ or $f^{rel}_{esc}<2\%$, due to the
exponential cut-off of the luminosity function at the bright end.

We have explored in Fig. \ref{evolfesc} five different parameterizations
for $f^{rel}_{esc}(L)$:
\begin{itemize}
\item
A: constant $f^{rel}_{esc}$ at 1.7\% at all luminosities (continuous red line).
This is not consistent with the limit of \cite{bayliss14}, but can
be easily scaled to $f^{rel}_{esc}\le 1\%$ with a simple
multiplicative factor.
\item
B: two-step function: 1.7\% at $M_{1500}\le -20.0$ and 10.3\% at fainter
magnitudes (long-dashed green line).
\item
C: three step function: 1.7\% at $M_{1500}\le -20.0$, 10.3\% at$-20.0<M_{1500}\le
-19.0$, and 23.2\% at fainter magnitudes (short-dashed grey line).
\item
D: exponential law $f^{rel}_{esc}=f_1*exp(-\alpha\cdot L/L*)$, with
$f_1=0.256$ and $\alpha=2.7$ (dotted magenta line).
\item
E: power law $f^{rel}_{esc}=f_1\cdot (L/L*)^{-\alpha}$
with $f_1=0.0167$ and $\alpha=0.79$ (dot-dashed dark green line).
\end{itemize}

The various evolutionary parameterizations of the escape fraction as a
function of the galaxy luminosity will be used in the next section to
derive different estimates of the average galaxy emissivity at $z\sim
3.3$.

\section{Estimating the Ionizing Emissivity produced
by star-forming galaxies at $z\sim 3$}

The rate of ionizing photons produced by star-forming galaxies can be
expressed as $\dot{n}_{ion}=\rho_{UV} \cdot \xi_{ion} \cdot
f^{abs}_{esc}$, where the three terms on the right side can be
evaluated upon reasonable assumptions.

The first term, $\rho_{UV}$, is the emissivity of high-z star-forming
galaxies, measured at a non-ionizing wavelengths, typically at 1500
{\AA} rest frame. It can be obtained by integrating the luminosity
function of galaxies, multiplied by their luminosities, down to faint
limits, e.g. $L\sim 10^{-3}-10^{-2}L^*$. The ionization efficiency
$\xi_{ion}$ measures the rate of ionizing photons produced by a given
amount of UV (1500 {\AA} rest frame) luminosity. It depends on the
nature of the stellar populations that developed in distant galaxies
(i.e. IMF, age, metallicity, binary fraction). Finally, the absolute
escape fraction $f^{abs}_{esc}$ measures the percentage of ionizing
radiation, produced by stars, which is not absorbed by gas and dust in
the ISM and is finally emitted in the surrounding IGM. In practice, it
is the ratio between the observed and produced fluxes at 900 {\AA}
rest frame. Numerous attempts have been carried out recently to
evaluate the ionization efficiency $\xi_{ion}$ of high-z galaxies
(\cite{dunlop13,robertson13,bouwens15ion} just to mention
several examples).

However, a number of assumptions are hidden on the calculations of
the rate of ionizing photons
$\dot{n}_{ion}$, and its derivation suffers from intrinsic
degeneracies between the physical properties assumed for the
underlying stellar populations.
More importantly, the two quantities $\xi_{ion}$ and
$f^{abs}_{esc}$ are by definition degenerate, since different
combinations of absolute escape fraction and production efficiency of
ionizing photons can give the same observed properties for a galaxy.
It is not easy, indeed, to provide an unbiased
estimation for the absolute escape fraction $f^{abs}_{esc}$. Since the
intrinsic SEDs of star-forming galaxies are not known a priori in the
far UV, assumptions on the physical properties of the galaxies
(stellar populations, age, star formation history, metallicity, etc.)
must be adopted. In particular, the choice of stellar population
synthesis model can change the intrinsic LyC emissivity up to a factor
of 1.5 or 2, if prescriptions for the treatment of the stellar
binaries are adopted (e.g. \cite{bpass,stanway}). Moreover, since the
age of our galaxies are not well constrained, the intrinsic LyC
production is uncertain by a similar factor.

We choose, instead, an alternative approach based on the relative
escape fraction, as described in Equation \ref{eqnfesc}. Following
\cite{grazian16}, we choose to derive the ionizing emissivity adopting
the following relation:

\begin{equation}
\rho_{900}^{esc}=\rho_{1500}\cdot f^{rel}_{esc}\cdot
(L_{900}/L_{1500})_{int} \, ,
\label{eqnemis}
\end{equation}
\noindent
where the factor $(L_{900}/L_{1500})_{int}$ is proportional to
$\xi_{ion}$. In this case the
advantage, thanks to Equation \ref{eqnfesc}, is that
$f^{rel}_{esc}\cdot (L_{900}/L_{1500})_{int}$ can be written as
$(F_U/F_R)_{obs}exp(\tau_{IGM})$ which is the product of two
quantities less subject to systematic uncertainties or assumptions,
and, neglecting the extinction by dust, it is equivalent to
$\xi_{ion} \cdot f^{abs}_{esc}$.

The ``relative'' escape fraction, by definition, depends on the
assumed ratio $(L_{1500}/L_{900})_{int}$ and, as such, it is not free
from assumptions on the physical properties of the galaxies. But its
main advantage, as discussed in \cite{grazian16}, is that its
dependencies on the physical properties of the galaxies are canceled
out when one computes the emissivity at $\lambda \le 900$ {\AA} rest,
by combining Eq.\ref{eqnfesc} with Eq.\ref{eqnemis}.
The only parameter critically affecting Eq.\ref{eqnemis} is the input
luminosity density of non-ionizing radiation, $\rho_{1500}$. To obtain
this value it is sufficient to integrate an observed luminosity function,
multiplied by the absolute luminosity, down to a specific magnitude limit.

In addition, it is useful to express the LyC emissivity $\rho_{900}^{esc}$
in terms of the \ion{H}{I} photo-ionizing background $\Gamma_{-12}$, in units of
$10^{-12}$ photons per second. Following \cite{grazian16}, we
have:

\begin{equation}
\Gamma_{-12}=\frac{10^{12} \cdot \rho_{900}^{esc} \cdot \sigma_{\ion{H}{I}} \cdot
\Delta l \cdot (1+z)^3}{h_P \cdot (3+|\alpha_{UV}|)} \, ,
\end{equation}
\noindent
where $\sigma_{\ion{H}{I}}=6.3\times 10^{-18} cm^2$ is the photo-ionization
cross section of hydrogen at $\lambda=912$ {\AA}, $\alpha_{UV}=-1.8$ is
the spectral slope of the stellar ionizing radiation (see \cite{fg09})
and $h_P$ is the
Planck constant. The mean free path of ionizing photons $\Delta l$ is
83.5 proper Mpc at z=3.3, following the fitting formula by
\cite{worseck14a}.

We derive the relevant quantities needed to compute $\Gamma_{-12}$
in the following sub-sections.

\subsection{The luminosity function of galaxies at $z\sim 3$}

\begin{figure}
\includegraphics[width=9cm,angle=0]{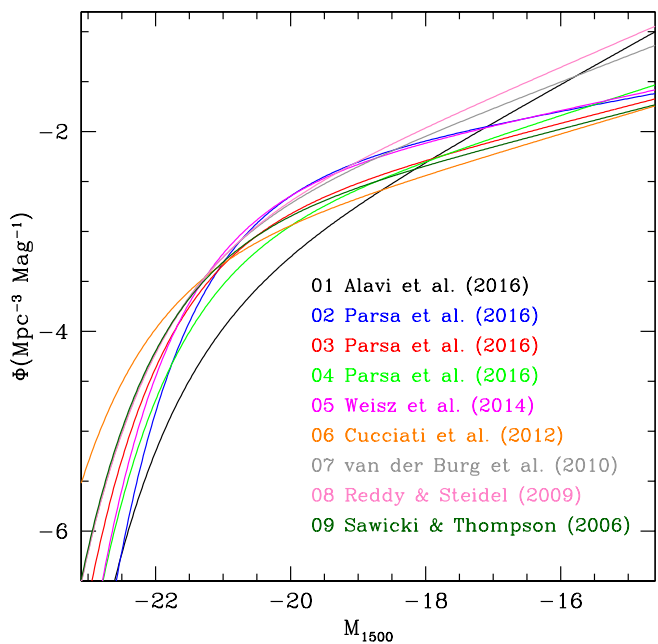}
\caption{
The galaxy luminosity functions at $z\sim 3$ from different work, as
summarized in Table \ref{table:lf}.
}
\label{summaryLF}
\end{figure}

An accurate estimate of the shape of the luminosity function at the
faint end is crucial to quantify the luminosity density $\rho_{1500}$,
and hence the \ion{H}{I} photo-ionization rate produced by SFGs at
$z\sim 3$. As in \cite{grazian16}, we have explored different
determinations for the luminosity function of star-forming galaxies at
these redshifts. We have collected relevant luminosity functions at
$z\sim 3$ from the literature, exploring the last ten years. In Table
\ref{table:lf} we have summarized the adopted parameters for the
Schechter function ($M^*$, $\Phi^*$, $\alpha$) of those recent
works. Fig. \ref{summaryLF} shows the different shapes of the galaxy
luminosity functions at $z\sim 3$.

\begin{table*}
\caption{The Schechter coefficients of the luminosity function
of star-forming galaxies at $z\sim 3$}
\label{table:lf}
\centering
\begin{tabular}{c c c c c c}
\hline
\hline
LF &  $M^*$ & $\Phi^*$ & $\alpha$ & Reference & Notes \\
\hline
01 & -20.71 & 0.00055 & -1.94 & Alavi et al. (2016) & LF at 2.2<z<3.0
1500 {\AA} rest frame \\
02 & -20.20 & 0.00532 & -1.31 & Parsa et al. (2016) & LF at z=2.8
1500 {\AA} rest frame \\
03 & -20.71 & 0.00206 & -1.43 & Parsa et al. (2016) & LF at z=3.8
1500 {\AA} rest frame \\
04 & -20.61 & 0.00161 & -1.54 & Parsa et al. (2016) & best fit at z=3.3
1500 {\AA} rest frame \\
05 & -20.45 & 0.00410 & -1.36 & Weisz et al. (2014) & LF z=3.0
1500 {\AA} rest frame \\
06 & -21.40 & 0.00086 & -1.50 & Cucciati et. al. (2012) &
1500 {\AA} rest frame \\
07 & -20.94 & 0.00179 & -1.65 & van der Burg et al. (2010) & z=3.0
1600 {\AA} rest frame \\
08 & -20.97 & 0.00171 & -1.73 & Reddy \& Steidel (2009) & z=3.05
1700 {\AA} rest frame \\
09 & -20.90 & 0.00167 & -1.43 & Sawicki \& Thompson (2006) & z=3.0
1700 {\AA} rest frame \\
\hline
\end{tabular}
\\
Cucciati et. al. (2012) didn't actually measure the faint end slope at
$z>1.7$, but fixed this value based on lower-z luminosity functions.
\end{table*}

A recent collection of luminosity functions from the literature has
been provided by \cite{parsa16}. They computed the luminosity function
in two redshift
bins, $z=2.8$ and $z=3.8$. Moreover, they also provide an analytic
formula to derive the parameters of the Schechter function depending
on redshift. In Table \ref{table:lf} we consider the best fit of the
relations given by \cite{parsa16} at $z=3.3$. The steep luminosity
functions by van der Burg et al. (2010) and \cite{rs09} are not
in agreement with the flatter ($\alpha\sim -1.4$) functions at $z\sim
3$ (Parsa et al. 2016). A notable exception is the result of
\cite{alavi16}, with a slope of -1.9 and extending down to
$M_{1500}=-12.5$, thanks to strong lensing magnification. It is
interesting to note that the luminosity functions
derived on lensing fields are usually
steeper than the one obtained in blank fields (see also
\cite{livermore16} and \cite{bouwens16} for similar comparisons at
$z\sim 6$). This fact could be related to possible, still unknown,
biases linked to the lensing amplification or completeness
corrections. However, the main aim here is not to do a critical review
of all the luminosity functions in the literature,
but to select a representative
sample in order to explore the implications on the calculations of the
photo-ionizing background.

From Fig. \ref{summaryLF} we can notice a large variance in the galaxy
number density, especially at faint luminosities.
At luminosities brighter than $0.5L^*$
($M_{1500}\le -20.2$) this scatter does not influence the estimate of the
photo-ionization rate $\Gamma_{-12}$: as already shown by
\cite{grazian16}, the contribution of bright galaxies is not relevant,
irrespective of the adopted parameterization of the luminosity
function. This is mainly due to the relatively low escape fraction
($\lesssim 2\%$) and to the low number densities
at these bright luminosities. At fainter magnitudes, we
can identify in Fig. \ref{summaryLF} three main families of luminosity
functions: 1-high space densities both at intermediate and faint
magnitudes, with $\Phi\sim 5\cdot 10^{-3} Mpc^{-3} Mag^{-1}$ at $M_{1500}=-19$ and
$\Phi\sim 4\cdot 10^{-2}$ at $M_{1500}=-16$ (07-van der Burg et al. 2010,
08-Reddy \& Steidel 2009); 2-high space density at intermediate
magnitudes but low density at fainter magnitudes, with $\Phi\sim
5\cdot 10^{-3}$ at $M_{1500}=-19$ and $\Phi\sim 10^{-2}$ at
$M_{1500}=-16$ (02-Parsa et al. 2016 z=2.8, 05-Weisz et al. 2014); 3-low space
densities both at intermediate and faint magnitudes, with $\Phi\sim
2\cdot 10^{-3}$ at $M_{1500}=-19$ and $\Phi\sim 10^{-2}$ at
$M_{1500}=-16$ (03-Parsa et al. 2016 z=3.8, 04-Parsa et al. 2016 best
fit at z=3.3, 06-Cucciati et al. 2012, 09-Sawicki \& Thompson 2006).
The luminosity function of Alavi et al. (2016) is an
exception in these three groups, with very low space density around
$L^*$, but a steep faint end ($\alpha\sim -1.9$).

In the next section we will use these luminosity functions to evaluate
the photo-ionization rate for different evolving escape fraction with
luminosities $f^{rel}_{esc}(L)$.

\subsection{Estimating the \ion{H}{I} photo-ionization rate}

Table \ref{table:gamma} summarizes the resulting photo-ionization rate
$\Gamma_{-12}$ for different parameterizations of $f^{rel}_{esc}(L)$
and for different luminosity functions (Models A-E and luminosity
functions 01-09,
respectively). The relevant emissivities have been computed integrating
the luminosity function from $M_{1500}=-26.0$ to $M_{1500}=-17.4$ mag,
corresponding to luminosities $4\times 10^{-2}L^* \le L\le 10^2L^*$.
The lower bound in luminosity has been set to the faintest lensed galaxy
for which we have an upper limit to the LyC escape fraction, the one
by \cite{amorin14}. We provide in Table
\ref{table:gamma} also the $\Gamma_{-12}$ obtained extending the lower limit
to $M_{1500}=-16.0$ mag, corresponding to a luminosity $L=10^{-2}L^*$.
For comparison, a recent estimate of the \ion{H}{I} photo-ionization rate of
$\Gamma_{-12}=0.79^{+0.28}_{-0.19}$ at z=3.2 has been provided by
\cite{bb13}, hereafter BB13. We should take into account the contribution
of the AGN population to $\Gamma_{-12}$: a conservative estimate
by Haardt \& Madau (2012) indicates
$\Gamma_{-12}=0.37$ at $z\sim 3$ (see their Fig. 8a). Similarly,
\cite{cristiani16} find a similar value $\Gamma_{-12}=0.36$ at $z=3.3$.
Thus, a contribution by galaxies of at least $\Gamma_{-12}\sim 0.4$
is required in order to explain the measured UVB.

\begin{table*}
\caption{The \ion{H}{I} photo-ionization rate $\Gamma_{-12}$ produced
by galaxies for different
functional forms of $f^{rel}_{esc}(L)$ and for different
parameterizations of the luminosity functions}
\label{table:gamma}
\centering
\begin{tabular}{c | c c c c c | c c c c c}
\hline
\hline
Models & A & B & C & D & E & A & B & C & D & E \\
\hline
LF & \multicolumn{5}{c}{$\Gamma_{-12}(-26.0\le M_{1500}\le -17.4)$} &
\multicolumn{5}{c}{$\Gamma_{-12}(-26.0\le M_{1500}\le -16.0)$} \\
\hline
01 & 0.04 & 0.22 & 0.39 & 0.34 & 0.19 & 0.06 & 0.34 & 0.65 & 0.62 & 0.56 \\
02 & 0.12 & 0.54 & 0.86 & 0.55 & 0.27 & 0.14 & 0.64 & 1.08 & 0.76 & 0.47 \\
03 & 0.09 & 0.35 & 0.55 & 0.49 & 0.26 & 0.10 & 0.42 & 0.71 & 0.65 & 0.47 \\
04 & 0.07 & 0.30 & 0.49 & 0.41 & 0.21 & 0.08 & 0.37 & 0.66 & 0.58 & 0.42 \\
05 & 0.13 & 0.53 & 0.83 & 0.63 & 0.32 & 0.15 & 0.63 & 1.05 & 0.85 & 0.56 \\
06 & 0.08 & 0.27 & 0.41 & 0.53 & 0.32 & 0.09 & 0.32 & 0.53 & 0.65 & 0.58 \\
07 & 0.13 & 0.52 & 0.86 & 0.86 & 0.48 & 0.15 & 0.68 & 1.20 & 1.21 & 1.04 \\
08 & 0.14 & 0.59 & 0.99 & 1.00 & 0.57 & 0.17 & 0.80 & 1.45 & 1.48 & 1.33 \\
09 & 0.09 & 0.33 & 0.50 & 0.50 & 0.27 & 0.10 & 0.39 & 0.64 & 0.64 & 0.48 \\
\hline
Average & 0.10 & 0.41 & 0.65 & 0.59 & 0.32 & 0.12 & 0.51 & 0.89 & 0.83 & 0.66 \\
\hline
\hline
\end{tabular}
\end{table*}

In order to interpret Table \ref{table:gamma}, we can divide the
Models A-E in three categories: the first one includes Model A
(constant escape fraction of 1.7\% at all luminosities), the second
group (Model B and C) is characterized by low $f^{rel}_{esc}\sim 1-2\%$
for galaxies brighter than $M_{1500}\sim -20$ and $f^{rel}_{esc}\sim 10-20\%$ for
fainter objects. Finally, a third group (Model D and E) has
$f^{rel}_{esc}\gtrsim 20\%$ fainter than $M_{1500}\sim -18$, and small
values $f^{rel}_{esc}\le 10\%$ brighter than $M_{1500}\sim -19$. As mentioned
above, the luminosity functions can be divided as well in three
main families, according to the space densities at $M_{1500}\sim -19$
and $M_{1500}\sim -16$.

Looking at the left side of Table \ref{table:gamma} (i.e. considering only
galaxies brighter than $M_{1500}=-17.4$), it is easy to conclude that Model
A has $\Gamma_{-12}\sim 0.04-0.14$, much lower than the required UVB value
of 0.4, for all the luminosity functions. Assuming a two-step
function for $f^{rel}_{esc}(L)$ (Model B), only the luminosity functions with
high space density at $M_{1500}\sim -19$ can provide enough \ion{H}{I} ionizing
photons at $z\sim 3$. Models C and D are instead able
to reach the required UVB for all the luminosity functions taken into
account. If the LyC escape fraction raises exponentially at lower
luminosities (Model E), then only the luminosity functions of
van der Burg et al. (2010, LF=07) and Reddy \& Steidel (2009, LF=08), with high
space densities both at $M_{1500}\sim -19$ and $M_{1500}\sim -16$,
are able to reach the critical value for $\Gamma_{-12}$.

It is interesting to note that Model C has the maximum contribution to
$\Gamma_{-12}$, w.r.t. the considered models A-E. Under this scenario,
galaxies at $M_{1500}\sim -19$ could be the main contributors to the
ionizing UVB and are crucial for our understanding of the
role of star formation on the reionization and post-reionization
phases.

The results above are valid if we take into account galaxies brighter
than $M_{1500}=-17.4$, or equivalently $L=0.036L^*$. If we extend the
lower bound to $M_{1500}=-16.0$ (right side of Table \ref{table:gamma})
or even fainter,
we obtain larger values of $\Gamma_{-12}$. In particular, all the considered
luminosity functions are always able to reach the required UVB,
except for constant escape fraction at the level of $\sim 1-2\%$
(Model A). Models with rapid increase of $f^{rel}_{esc}$ at faint
luminosities (Models B, C, D, and E) give $\Gamma_{-12}$ consistent with
BB13, or even a factor of 2-3 higher when they are integrated down to
$M_{1500}=-13.5$. In this case, extremely faint galaxies,
with luminosities between $10^{-3}L^*$ and $10^{-2}L^*$ are important for
reproducing the UVB at $z\sim 3$, provided that the escape fraction
is larger than 2\% for all the faint star-forming galaxies. These
regimes of luminosities will be hopefully investigated thanks to
strong lensing magnification in the future (e.g. \cite{vanzella12b}).


\section{Discussion}

\subsection{Comparison with predictions by theoretical models}

It is interesting at this stage to compare the observed limits for
$f^{rel}_{esc}$ shown in Fig. \ref{evolfesc} with predictions from
theoretical models. In general, many theoretical models predict
ionizing escape fractions increasing for progressively fainter
galaxies. A recent exception to this scenario is provided by the
\cite{sharma16b} model where the escape fraction increases with the
star formation surface density $\Sigma_{SF}$ on scales of 1 kpc
reaching 5\% and 13\% in galaxies with $M_{1500}=-21.4$ and $-23.9$,
respectively. These predicted values are in contrast with our limits,
which have been
observed in galaxies in the same luminosity interval. We note that the
starburst galaxy cited by \cite{sharma16b} as a clear
prototypical example of luminous LyC emitter shows X-ray
detection probably associated with AGN activity (see Guaita et al. 2016).

Price et al. (2016) and Faisst (2016) derived empirical constraints
from low-redshift observations in order to get constraints on the
escape fraction and LyC production efficiency of galaxies. They assume
a redshift evolution of the absolute escape fraction starting at z=3.3
from values in the range $f^{abs}_{esc}=3.5-9\%$, in tension with our
observed value derived for bright galaxies ($M_{1500}\le -20.0$).
Adopting values of 1-2\%, in better agreement with our
limits, would imply a rapid evolution of the escape fraction with a
quick increase up to the maximum redshift $z=8-9$ allowed by the late
reionization scenario suggested by the Planck 2016 data. It is also
interesting to note that the empirical relation between [OIII]/[OII] ratio
and $f^{abs}_{esc}$ adopted by Faisst (2016, Fig. 2) is not consistent
with the observations of \cite{christensen12,amorin14,bayliss14}. At
$z>3$ and at relatively faint magnitudes, indeed, the latter have
found that $f^{rel}_{esc}\le 1\%$ and 23\% at $M_{1500}=-20.5$ and
$M_{1500}=-17.4$, respectively, while the observed line ratio is
[OIII]/[OII]$>10$. It is thus important to note that a causal
connection between the high ionization
efficiency and the large escape fraction of ionizing radiation has
not been established yet.

Anderson et al. (2016), Kimm et al. (2016), Gnedin (2016),
Hassan et al. (2016), Xu et al. (2016), and Yoshura et al. (2016)
have used numerical
cosmological simulations and theoretical tools to analyze the
contribution of faint galaxies to the reionization process. From these
studies it appears that the contribution to the reionization by
star-forming galaxies does not strongly depend on redshift. For this
reason, their estimates can be compared to the lower average redshift
$z\sim 3.3$ of our sample. The predicted escape fractions are in
general constant or increasing for progressively fainter galaxies,
with values reaching up to $40-60$\% but keeping as low as few percent
for the brighter fraction ($M_{1500}<-20$).

In summary, all the recent theoretical models, with the exception of
\cite{sharma16a} and \cite{sharma16b}, tend to agree on relatively
small values, of the order of few percent, for the escape fraction at
$z\sim 3$ and at luminosities close to $L^*$. They are thus in
agreement with the observational results summarized in
Fig. \ref{evolfesc}.

\subsection{Comparison with recent works on LyC escape fraction of
star-forming galaxies}

In this sub-section we compare our results, summarized in
Fig. \ref{evolfesc} and Table \ref{table:fesc}, with recent
measurements of the escape fraction of star-forming galaxies at
high-z. We have divided the literature papers in two main groups, the
first containing results based on the indirect signatures of possible
LyC emission, while the second one summarizes the evidences of direct LyC
detections.

\subsubsection{Indirect LyC measurements with other proxies}

The long search for LyC emitters during the last 15 years has
stimulated the investigation of indirect proxies to pinpoint robust
samples of \ion{H}{I} ionizers both in the local Universe and at high
redshift. An efficient method, based on high [OIII]/[OII] line ratio or
other high excitation lines, has been proposed by a number of works
(\cite{nakajimaouchi,stark15,schaerer16}, just to provide an example) and
has been supported by the results of \cite{izotov15,izotov16}, and
\cite{vanzella16}.

Usually, the high ionization efficiency of LyC emitter candidates is
also associated with compact morphology and possibly strong outflows.
Recently, Nakajima et al. (2016), Trainor et al. (2016), and Faisst (2016) have
proposed to search for LyC emitters in galaxies with high ionizing
ratio, possibly due to the low metallicity or to the
harder ionization spectrum (e.g. by a top-heavy initial mass
function). Reddy et al. (2016) and Leethochawalit et al. (2016), instead,
suggested to search for LyC
emitters among the sources with low covering fraction of neutral
hydrogen, where the ISM is porous and has holes (the ionization-bound
nebula of Zackrisson et al. 2013).

While all these methods are promising, it is worth stressing that they
are however far from being robust indicators of appreciable LyC
emission. Just to recall two notable examples discussed above, the galaxies by
\cite{amorin14} and by \cite{vasei16} showed high [OIII]/[OII] line ratio
coupled with upper limits on LyC escape fraction at $\sim 10-20\%$
level, obtained through direct measurements.
Moreover, \cite{rutkowski16} find that strong H-$\alpha$ emitters
at $z\sim 1$, which are thought to be close analogs of high redshift
sources of reionization, show an escape fraction $\le 9.6\%$ (at $3\sigma$),
again through direct LyC measurements.
This suggests the need for more accurate calibrations of the
indirect methods with direct LyC measurements.

\subsubsection{Direct LyC measurements}

Micheva et al. (2017) presented a sample of 18 LAEs and 7 LBGs
detected in LyC out of a sample of 159 LAEs and 136 LBGs in the SSA22
field. The relative escape fraction for the LyC detected LAEs is
$\sim30\%$, while for the LyC detected LBGs it is $\sim20\%$. However,
a great fraction of the LyC candidates shows spatial offsets between
the rest-frame non-ionizing detection and the LyC-emitting
substructure or between the Lyman-$\alpha$ emission and the \ion{H}{I}
ionizing radiation casting doubts on the fraction of contamination by
lower redshift interlopers.

\cite{smith16} have used the Early Release Science (ERS) data of
the UV sensitive WFC3 instrument onboard HST to measure the LyC escape
fraction for a sample of 50 massive SFGs at $z\sim 2.3-5.8$. They
reported a $>3\sigma$ detection of $f^{abs}_{esc}\sim 0.1\%$ at $z\sim
2.4-2.7$ and a marginal $\le 3\sigma$ LyC signal at $z\sim 3$.
These data do not show any significant correlation of $f^{abs}_{esc}$ with
the galaxy luminosity.

A different approach has been recently adopted by Shapley et
al. (2016). They started from deep Keck/LRIS UV spectroscopy for a
large sample of LBGs at $z\sim 3$, as described by
\cite{reddy16}. Roughly 10\% of their galaxies show detections of LyC
radiation (Steidel et al. in prep.). In particular, Shapley et
al. (2016) studied the case of an intriguing spectroscopic LyC
detection at $z\sim 3.2$ corroborated by HST data, which allows them
to exclude contamination by foreground objects. This galaxy has
$f^{abs}_{esc}\sim 42\%$, but also shows evidence of high column
density of neutral hydrogen, since the Lyman-$\beta$ and
Lyman-$\gamma$ lines in absorption are saturated. Moreover, the age of
this galaxy (1.3 Gyr) is not as young as expected for strong LyC
emitters, which should have young stellar populations of several Myr
to provide the required ionizing photon efficiency.

Recently, starting from the stack spectrum of 33 galaxies at $z\sim 4$
from the VUDS survey, \cite{marchi16} have derived a $\sim 2\sigma$
detection of LyC radiation, measuring $f^{rel}_{esc}=0.09\pm
0.04$. They also identified a possible LyC emitting galaxy at $z\sim
3.6$ with high EW in Lyman-$\alpha$. These results are in agreement
with previous results (Vanzella et al. 2010, Boutsia et al. 2011,
Grazian et al. 2016, Guaita et al. 2016, Japelj et al. 2017)
and with our findings that the contribution of the whole galaxy population
to the ionizing background is probably modest, extending the study at
$z\sim 4$.

In summary, the few detections of LyC emission from SFGs at $z\ge 3$
indicate that they are rare events or that the distribution of
escape fraction is bimodal. Only a limited sub-sample (few
percent) of the distant galaxies presents significant ($\sim 20-40\%$)
escape fraction, while the general galaxy population usually does not
show large $f_{esc}$ of ionizing photons (but see Steidel et al. in prep.).
Our result, where no individual robust detection has been found out of
a sample of 69 SFGs at $z\sim 3.3$, is in agreement with the recent
achievements outlined above by Smith et al. (2016), Vanzella et al. (2016),
Shapley et al. (2016), Guaita et al. (2016), Micheva et al. (2017),
Japelj et al. (2017), Marchi et al. (2017). It is thus possible that the escape
of ionizing radiation is a peculiar phenomenon, associated with
particular phases of the galaxy lifetime or to the action of supernovae/AGN
feedback on their ISM.


\section{Summary and Conclusions}

We have used ultra-deep observations ($\gtrsim 30$ hours) of the
CANDELS/GOODS-North field with the LBC instrument in the U and R bands
with the aim of determining the LyC escape fraction for a sample of
faint ($L<0.5L^*$) star-forming galaxies at $z\sim 3.3$. To this aim,
we complemented the CANDELS/GOODS-North sample with additional galaxies
from the EGS and COSMOS fields, reaching a total of 69 star-forming
galaxies of $-23\lesssim M_{1500}\lesssim -19$ in the redshift interval
$3.27<z<3.40$. This paper is thus an extension of the work by
\cite{grazian16}, both on the statistics (almost doubling the sample) and
on the luminosity range analyzed.

Several ingredients are needed in order to reliably compute the LyC
escape fraction: 1) very deep images at 900 and 1500 {\AA} rest-frame,
in this case the U and the R bands at $z\sim 3$; 2) images with high
spatial resolution, possibly from HST in different filters, in
order to clean the sample from lower redshift contaminants; 3) high
quality spectroscopic redshifts in a particularly narrow redshift
range, in this case $3.27<z<3.40$, allowing the U filter of LBC to
sample only LyC photons close to 900 {\AA} rest-frame; 4) high number
statistics, in order to reduce the stochasticity of the IGM
absorption; 5) deep X-ray data, in order to exclude the AGNs
from our sample.

We have chosen the CANDELS/GOODS-North, EGS, and COSMOS
datasets since in all the three fields we have these ingredients, i.e.
deep LBC data in the U and in the R bands in wide areas,
HST coverage from CANDELS, AEGIS, and COSMOS, X-ray images from
Chandra, spectroscopic redshifts from the CANDELS, AEGIS, and
VUDS-COSMOS databases. In particular, the critical ingredient of this
work is the ultra-deep U-band image of LBC in the CANDELS/GOODS-North
field, covering an area of $\sim 0.6$ sq. deg. to a limit of 30.2 AB
mag (at 1 $\sigma$). This image allows us to put stringent limits to the
LyC escape fraction of faint star-forming galaxies at $z\sim 3$.

We have cleaned the $z\sim 3$ sample from the contamination of AGNs
and lower redshift interlopers, thanks to the Chandra and HST data.
We have used an efficient technique to compute the escape fraction
from deep LBC data, based on the PSF-matched photometry, as carried out
in \cite{grazian16}. This reduces the uncertainties on the
photometry and thus on the leakage of ionizing photons.

In order to explore the very faint regime, we added to our sample a
lensed galaxy studied by \cite{vanderwel13} and \cite{amorin14}. It is
a star forming galaxy in the COSMOS field at z=3.4 lensed by an
elliptical at $z\sim 1.5$. The upper limit in the U band by LBC allows
us to place an upper limit to the escape fraction of 23\% for an
object with luminosity of $0.036 L^*$. We added also other two lensed
galaxies from the literature (Christensen et al. 2012, Bayliss et al. 2014).
For the last two galaxies we do not have deep LBC images in the U band,
and the upper limit to the LyC escape fraction comes from the estimate of
the column density of neutral hydrogen from deep spectra in the
Lyman-$\alpha$ region. The lensed galaxies by
\cite{christensen12,amorin14} and \cite{bayliss14} are particularly
important since they are intrinsically faint (and low mass) objects
with high [OIII]/[OII] line ratios. They thus represent the interesting
sub-class of galaxies with high ionization efficiency, which are
supposed to emit copious amount of ionizing photons.

We derive an upper limit to the LyC $f^{rel}_{esc}$ for each
individual galaxy of the sample. We also stack the U and R bands of
our sample to derive a limit for the whole population of SFGs at
$z\sim 3$.
The main results are the following:

\begin{itemize}
\item
Out of 69 galaxies, we detect LyC emission only in 3 (two discussed in
\cite{grazian16} and the third one in the EGS field, ID=11372).
The source ID=16479 in the CANDELS/GOODS-North field has been discarded
since it is possibly affected by contamination.
\item
Stacking the U- and the R-band images of the whole sample, we derive an
upper limit of $f^{rel}_{esc}\le 1.7\%$ at S/N=1,
more stringent than previous results
(e.g. Chen et al. 2007, Vanzella et al. 2010, Boutsia et al. 2011,
Grazian et al. 2016, Guaita et al. 2016, Japelj et al. 2017,
Marchi et al. 2017).
\item
Dividing the total sample in different luminosity intervals, we
derive limits of $f^{rel}_{esc}\le 2.5\%$, 2.0\%, 2.4\%, and 10.3\%
at $M_{1500}=-22.8$, $M_{1500}=-21.7$, $M_{1500}=-20.8$, and
$M_{1500}=-19.7$, respectively.
The apparent increase of the escape fraction toward faint luminosities is
due to an observational limit in the depth of the U band.
\item
We have explored different dependencies of the relative escape fraction
on the galaxy luminosity, and provided different parameterizations to
reproduce the observed trends of $f^{rel}_{esc}(L)$.
\item
We have combined the derived parametric forms of $f^{rel}_{esc}(L)$
with different realizations of the galaxy luminosity function at
$z\sim 3$ taken from the literature,
and we have evaluated the emissivity at 900 {\AA} rest-frame, $\rho_{900}^{esc}$.
\item
Translating this quantity into a neutral hydrogen photo-ionization rate
$\Gamma_{-12}$,
we can conclude that $z\sim 3$ galaxies brighter than
$M_{1500}=-17.4$ have difficulties at
providing the UVB measured in the Lyman forest at $z\sim 3$,
especially if their LyC escape fraction is below 10\%.
\item
Fainter galaxies, with $M_{1500}\le -16$, can produce the observed UVB
if their LyC escape fraction is above 10\% at low luminosities.
If a fainter limit of $M_{1500}\sim -13$ is adopted,
then the constraints on $f^{rel}_{esc}$ can be relaxed.
\item
For two lensed galaxies with extreme line ratio of [OIII]/[OII]$>10$,
there are stringent upper limits of $f^{rel}_{esc}\le 1\%$
and 23\% at $M_{1500}=-20.5$ and $M_{1500}=-17.4$, respectively
(Bayliss et al. 2014, Amorin et al. 2014).
Any causal connection between the high ionization
efficiency and the large escape fraction of LyC radiation still awaits
confirmations by a larger sample.
\end{itemize}

Placing these results in a cosmological context, it is interesting to
note that if such low values of escape fraction will be found also at
$z>4$ and for faint luminosities, then star-forming galaxies would be
insufficient to cause the \ion{H}{I} reionization process.

Future work will include the derivation of a robust estimate
of the faint end of the luminosity function
at $z\sim 3$ in large areas thanks to the deep
U-band images in the COSMOS (Boutsia et al. 2014), EGS, and ultra-deep
CANDELS/GOODS-North data (\cite{grazLBC}). The combination of wide and
deep data by LBC, together with large spectroscopic databases
available in these fields, will allow us to measure the slope $\alpha$
with unprecedented precision. This is a crucial item (as well as the
measurements of the escape fraction) in the derivation of precise
contribution of faint galaxies to the UVB at z=3. The
CANDELS/GOODS-North sample can be used in the future also to confirm the
spectroscopic redshifts for other
star-forming galaxies at $3.27<z<3.40$ in the faint regime ($25.5<R<26.0$)
with deep optical spectroscopy.
At present, in fact, we are using in this work only 8
galaxies with spectroscopic redshifts available out of 48 (16\%)
galaxies at $3.27\le z_{phot}\le 3.40$ and $R\le 26.0$ in this
field. While this spectroscopic sub-sample is relatively small, it is
nonetheless representative for the faint star-forming galaxy
population at $z\sim 3$, indicating that our results are not biased by
low number statistics even at these faint limits.
The spectroscopic completeness of galaxies at $R\le 26.0$
can be enhanced in the future by adding $\sim$40 galaxies
and improving the limit of $f^{rel}_{esc}$ down to $\sim 4\%$.

In the future, it will be crucial to measure
the escape fraction at faint magnitudes, $M_{1500}\sim -16$ for a large
number of galaxies. At this
aim, strong lensing magnification (either through galaxy-galaxy
lensing or by massive clusters) will be a crucial instrument to
explore luminosity regimes which are otherwise inaccessible with the
present instrumentation. The rapid activities on this side (e.g. SLACS,
SL2S surveys for galaxy-galaxy lensing; HST Frontier Fields for
lensing by clusters) will give fresh data in the near future.

\begin{acknowledgements}
We warmly thank the anonymous referee for her/his useful suggestions and
constructive comments that help us to improve this paper.
We acknowledge financial contribution from the agreement ASI-INAF I/009/10/0.
The LBT is an international collaboration among institutions in the
United States, Italy, and Germany. LBT Corporation partners are The
University of Arizona on behalf of the Arizona university system;
Istituto Nazionale di Astrofisica, Italy; LBT
Beteiligungsgesellschaft, Germany, representing the Max-Planck
Society, the Astrophysical Institute Potsdam, and Heidelberg
University; The Ohio State University; and The Research Corporation,
on behalf of The University of Notre Dame, University of Minnesota,
and University of Virginia.
Based on observations collected at the European Organisation for Astronomical
Research in the Southern Hemisphere under ESO programme(s) 185.A-0791.
Based on observations obtained with MegaPrime/MegaCam, a joint project
of CFHT and CEA/IRFU, at the Canada-France-Hawaii Telescope (CFHT)
which is operated by the National Research Council (NRC) of Canada,
the Institut National des Science de l'Univers of the Centre National
de la Recherche Scientifique (CNRS) of France, and the University of
Hawaii. This work is based in part on data products produced at
Terapix available at the Canadian Astronomy Data Centre as part of the
Canada-France-Hawaii Telescope Legacy Survey, a collaborative project
of NRC and CNRS.
\end{acknowledgements}

%
%

\end{document}